\documentclass[twocolumn]{aastex63}
\usepackage{multirow}
\usepackage{comment}
\usepackage{lineno}
\usepackage{epsfig}
\usepackage{natbib}
\usepackage{amsmath}
\usepackage{hyperref}

\usepackage{xcolor}
\definecolor{forestgreen}{rgb}{0.13, 0.75, 0.13}

\newcommand{\zindex}{z}

\newcommand{\totsn}{1701}

 \newcommand{\numSDSS}{321}
 \newcommand{\numSNLS}{160}
 \newcommand{\numPS}{269}
 \newcommand{\numDES}{203}
 \newcommand{\numCSP}{89}
 \newcommand{\numCFAa}{13}
 \newcommand{\numCFAb}{24}

 \newcommand{\numCFAt}{92}
  \newcommand{\numLOWZ}{46}

 \newcommand{\numCFAf}{50}
 \newcommand{\numFOUND}{179}
 \newcommand{\numASASSN}{15}
 
 \newcommand{\numKAITm}{48}
 \newcommand{\numKAITs}{105}
 \newcommand{\numSWIFT}{57}
 
 \newcommand{\numHST}{16}
 \newcommand{\numSCP}{6}
 \newcommand{\numCANDELS}{8}

\begin{document}

\title{The Pantheon+ Analysis: The Full Dataset and Light-Curve Release}

\author{Dan Scolnic}
\affil{Department of Physics, Duke University, Durham, NC 27708, USA}
\email{daniel.scolnic@duke.edu}

\author{Dillon Brout}
\affil{Center for Astrophysics, Harvard \& Smithsonian, 60 Garden Street, Cambridge, MA 02138, USA}
\affil{NASA Einstein Fellow}
\email{dillon.brout@cfa.harvard.edu}

\author{Anthony Carr}
\affil{School of Mathematics and Physics, University of Queensland, Brisbane, QLD 4072, Australia}

\author{Adam G.\ Riess}
\affil{Space Telescope Science Institute, 3700 San Martin Drive, Baltimore, MD 21218, USA}
\affil{Department of Physics and Astronomy, Johns Hopkins University, Baltimore, MD 21218 USA}

\author{Tamara M.\ Davis}
\affil{School of Mathematics and Physics, University of Queensland, Brisbane, QLD 4072, Australia}

\author{Arianna Dwomoh}
\affil{Department of Physics, Duke University, Durham, NC 27708, USA}

\author{David O.\ Jones}
\affil{NASA Einstein Fellow}
\affil{Department of Astronomy and Astrophysics, University of California, Santa Cruz, CA 92064, USA}

\author{Noor Ali}
\affil{Umeå University, 901 87, Umeå, Sweden}

\author{Pranav Charvu}
\affil{Department of Physics, Duke University, Durham, NC 27708, USA}

\author{Rebecca Chen}
\affil{Department of Physics, Duke University, Durham, NC 27708, USA}

\author{Erik R.\ Peterson}
\affil{Department of Physics, Duke University, Durham, NC 27708, USA}

\author{Brodie Popovic}
\affil{Department of Physics, Duke University, Durham, NC 27708, USA}

\author{Benjamin M.\ Rose}
\affil{Department of Physics, Duke University, Durham, NC 27708, USA}

\author{Charlotte M. Wood}
\affil{Department of Physics and Astronomy, University of Notre Dame, Notre Dame, IN 46556, USA}

\author{Peter J.\ Brown}
\affil{Department of Physics and Astronomy, Texas A\&M University, 4242 TAMU, College Station, TX 77843, USA}
\affil{George P.\ and Cynthia Woods Mitchell Institute for Fundamental Physics \& Astronomy, College Station, TX 77843, USA}

\author{Ken Chambers}
\affil{Institute of Astronomy, University of Hawaii, 2680 Woodlawn Drive, Honolulu, HI 96822, USA}

\author{David A.\ Coulter}
\affil{Department of Astronomy and Astrophysics, University of California, Santa Cruz, CA 92064, USA}

\author{Kyle G. Dettman}
\affiliation{Department of Physics and Astronomy, Rutgers, the State University of New Jersey, Piscataway, NJ 08854, USA }

\author{Georgios Dimitriadis}
\affil{School of Physics, Trinity College Dublin, The University of Dublin, Dublin 2, Ireland}

\author{Alexei V.\ Filippenko}
\affil{Department of Astronomy, University of California, Berkeley, CA 94720-3411, USA}
\affil{Miller Institute for Basic Research in Science, University of California, Berkeley, CA 94720, USA}

\author{Ryan J.\ Foley}
\affil{Department of Astronomy and Astrophysics, University of California, Santa Cruz, CA 92064, USA}

\author{Saurabh W.\ Jha}
\affiliation{Department of Physics and Astronomy, Rutgers, the State University of New Jersey, Piscataway, NJ 08854, USA }

\author{Charles D.\ Kilpatrick}
\affil{Center for Interdisciplinary Exploration and Research in Astrophysics
(CIERA), Northwestern University, Evanston, IL 60208, USA}

\author{Robert P.\ Kirshner}
\affil{Center for Astrophysics, Harvard \& Smithsonian, 60 Garden Street, Cambridge, MA 02138, USA}
\affil{Gordon and Betty Moore Foundation, Palo Alto, CA 94304, USA}

\author{Yen-Chen Pan}
\affil{Graduate Institute of Astronomy, National Central University, 32001 Jhongli, Taiwan}

\author{Armin Rest}
\affil{Space Telescope Science Institute, Baltimore, MD 21218, USA}

\author{Cesar Rojas-Bravo}
\affil{Department of Astronomy and Astrophysics, University of California, Santa Cruz, CA 92064, USA}

\author{Matthew R.\ Siebert}
\affil{Department of Astronomy and Astrophysics, University of California, Santa Cruz, CA 92064, USA}

\author{Benjamin E.\ Stahl}
\affil{Department of Astronomy, University of California, Berkeley, CA 94720-3411, USA}

\author{WeiKang Zheng}
\affil{Department of Astronomy, University of California, Berkeley, CA 94720-3411, USA}

\submitjournal{Astrophysical Journal Letters}

\begin{abstract}
Here we present \totsn~light curves of 1550 unique, spectroscopically confirmed Type Ia supernovae (SNe~Ia) that will be used to infer cosmological parameters as part of the Pantheon+ SN analysis and the SH0ES (Supernovae and H$_0$ for the Equation of State of dark energy) distance-ladder analysis.  This effort is one part of a series of works that perform an extensive review of redshifts, peculiar velocities, photometric calibration, and intrinsic-scatter models of SNe~Ia.  The total number of light curves, which are compiled across 18 different surveys, is a significant increase from the first Pantheon analysis (1048 SNe), particularly at low redshift ($z$).  Furthermore, unlike in the Pantheon analysis, we include light curves for SNe with $z<0.01$ such that SN systematic covariance can be included in a joint measurement of the Hubble constant (H$_0$) and the dark energy equation-of-state parameter ($w$). We use the large sample to compare properties of 151 SNe~Ia observed by multiple surveys and 12 pairs/triplets of ``SN siblings'' --- SNe found in the same host galaxy.  Distance measurements, application of bias corrections, and inference of cosmological parameters are discussed in the companion paper by Brout et al. (2022b), and the determination of H$_0$ is discussed by Riess et al.~(2022).  These analyses will measure $w$ with $\sim 3$\% precision and H$_0$ with $\sim1$~km/s/Mpc precision.

\end{abstract}




\bigskip
\section{Introduction}

Measurements of Type Ia supernovae (SNe~Ia) were essential to the discovery of the accelerating expansion of the universe \citep{Riess1998,Perlmutter1999}. Since then, the continually growing sample size of these special ``standardizable candles'' has strengthened a key pillar of our understanding of the standard model of cosmology in which the universe is dominated by dark energy and dark matter.  While modern transient surveys are now discovering as many SNe~Ia in 5~yr as had been discovered in the last 40~yr (e.g., \citealp{Smith2020,dhawan21,jones21}), progress in using these data for constraining cosmological parameters has been made by the compilation of multiple samples (e.g., \citealp{Betoule2014,Scolnic2018,Brout2019,Jones19}).  The reason for this is that different surveys are optimized to discover and measure SNe in different redshift ranges, and the constraints on cosmological parameters benefit from leveraging measurements at different redshifts.  In this paper, we present the latest compilation of spectroscopically-confirmed SNe~Ia, which we call Pantheon+; this sample is a direct successor of the Pantheon analysis \citep{Scolnic2018}, which itself succeeded the Joint Light-curve Analysis \citep[JLA;][]{Betoule2014}.

In the past, measurements of the equation-of-state parameter of dark energy ($w$) and the expansion rate of the universe (H$_0$) have been done separately (e.g., \citealp{Riess2016}; \citealp{Scolnic2018}), even though both rely on many of the same SNe~Ia.  One reason for this split is that the determination of these two parameters is based on comparing SNe~Ia  in different redshift ranges.  For H$_0$, SNe~Ia in very nearby galaxies with $z \lesssim 0.01$ that have calibrated distance measurements are compared to those in the ``Hubble flow'' at $0.023<z<0.15$, ignoring higher redshifts.  For $w$, measurements typically  utilize SNe~Ia up to $z\approx2$, but exclude those at $z<0.01$. Thus, only SNe~Ia within one of the three ranges, those at $0.023 < z < 0.15$, are common to both analyses.

Here we perform a single analysis of discovered SNe~Ia measured over the entire redshift range, from $z=0$ to $z=2.3$.  This work spawns a number of analyses which include the $w$ measurement presented by Brout et al. (2022b, in prep., hereafter B22b) as well as the H$_0$ measurement of Riess et al. (2022, in prep., hereafter R22).  R22 additionally depend on Cepheid and geometric distance measurements, which make up what is called the ``first rung'' of the distance ladder, whereas Cepheid measurements and $z<0.01$ SN measurements make up the ``second rung,'' and SN measurements along with their redshifts make up the ``third rung.''  Both Cepheids and SNe are used in two of the three rungs.  Furthermore, the SNe discussed here can be used to measure growth of structure, as indicated by the model comparisons by \cite{Peterson2021} and for measurements of anisotropy discussed by B22b.  A review of many potential cosmological measurements possible with large SN~Ia samples is given by \cite{Scolnic_Decadal}.

Measurements of SN~Ia light curves by different surveys can be accumulated to improve their constraining power on cosmological inferences because (1) the SNe can be uniformly standardized using their light-curve shapes and colors, and any dependence of the standardization properties with redshift can be measured; and (2) properties of the photometric systems and observations of tertiary standards are typically given so that current analyses can recalibrate the systems (e.g., \citealp{Scolnic2015, Currie20}) and refit light curves.  This latter point, when used with an analysis of SN surveys in aggregate, yields the ability to quantify and reduce survey-to-survey calibration errors.  This is explored by Brout et al. (2022a, in prep., hereafter B22a), who present a new cross-calibration of the photometric systems used in this analysis and the resulting recalibration of the SALT2 light-curve model.  \cite{Brownsberger21} show that while measurements of H$_0$ are particularly robust to calibration errors of SNe~Ia, this is not the case for measurements of $w$.  In this paper, we analyze measurements of the same SNe from different surveys as an alternate test on the accuracy of our calibration.

The large size of this sample also allows us to compare ``sibling SNe'' --- that is, SNe belonging to the same host galaxy.  As shown in various studies \citep{Scolnic20,Burns20,Biswas21}, sibling SNe provide powerful tests of our understanding of the relationships between SN properties and their host galaxies.  With this large compilation, we can increase the statistics of sibling pairs (and triples).  Our findings on the consistency of the distance modulus values determined for sibling SNe, as well as the consistency of distance measurements of SNe from different samples, can be used to improve the construction of the distance-covariance matrix between SNe.  This matrix is described by B22b, and relates the covariance between distance measurements of SNe due to various systematic uncertainties.

Lastly, this paper documents the data release of standardized SNe~Ia for the Pantheon+ sample.  A companion paper by \cite[][hereafter C22]{carr2021pantheon} performs a comprehensive review of all the redshifts used and also corrects a small number of SNe with incorrect meta-properties (e.g., location, host association, naming),  all included here.  We note that this compilation includes light curves that have not been published elsewhere and light curves that have been provided individually as the focus of a single paper, as well as the larger samples from specific surveys. The compilation presented here attempts to homogenize the presentation and documentation of these light curves. 

The structure of this paper is as follows. In Section 2, we describe the light-curve samples released as part of the Pantheon+ compilation. Section 3 presents the light-curve fits, the selection requirements (data quality cuts), and the properties of the host galaxies.  We discuss in Section 4 trends of the fitted and host-galaxy parameters, as well as new studies of SN siblings and duplicate SNe.  Section 5 presents our discussions and conclusions. Importantly, in the Appendix, we describe the format of the data release itself.

\section{Data}

\begin{figure}[htp]

\includegraphics[clip,width=\columnwidth]{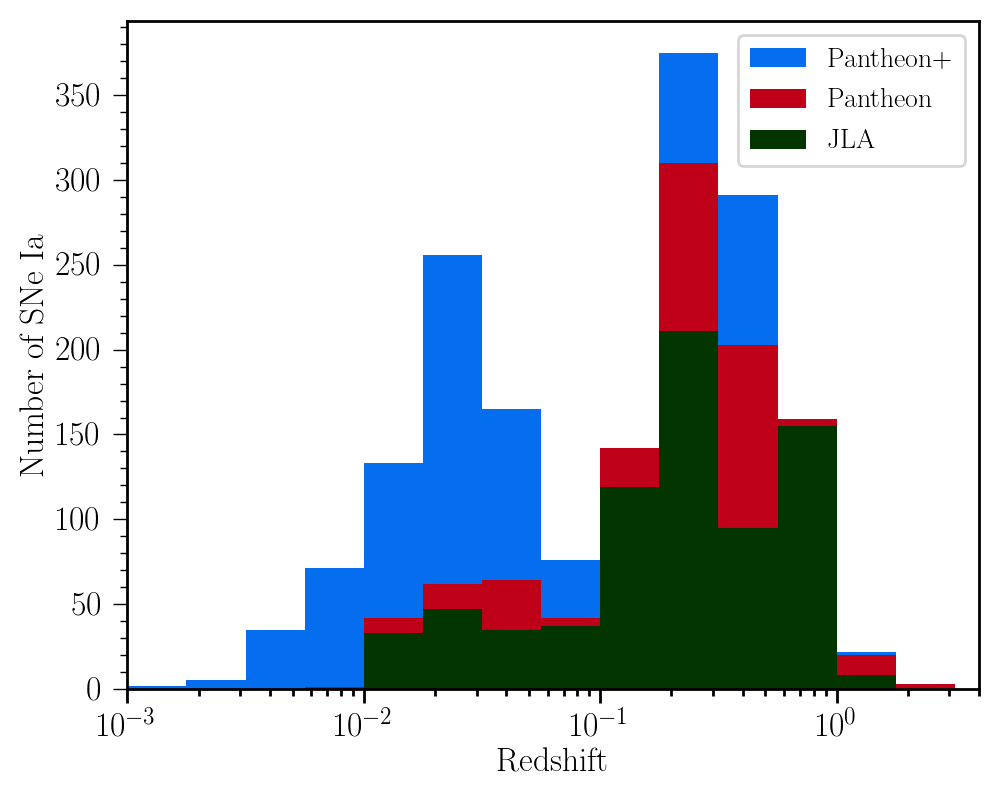}%

\includegraphics[clip,width=\columnwidth]{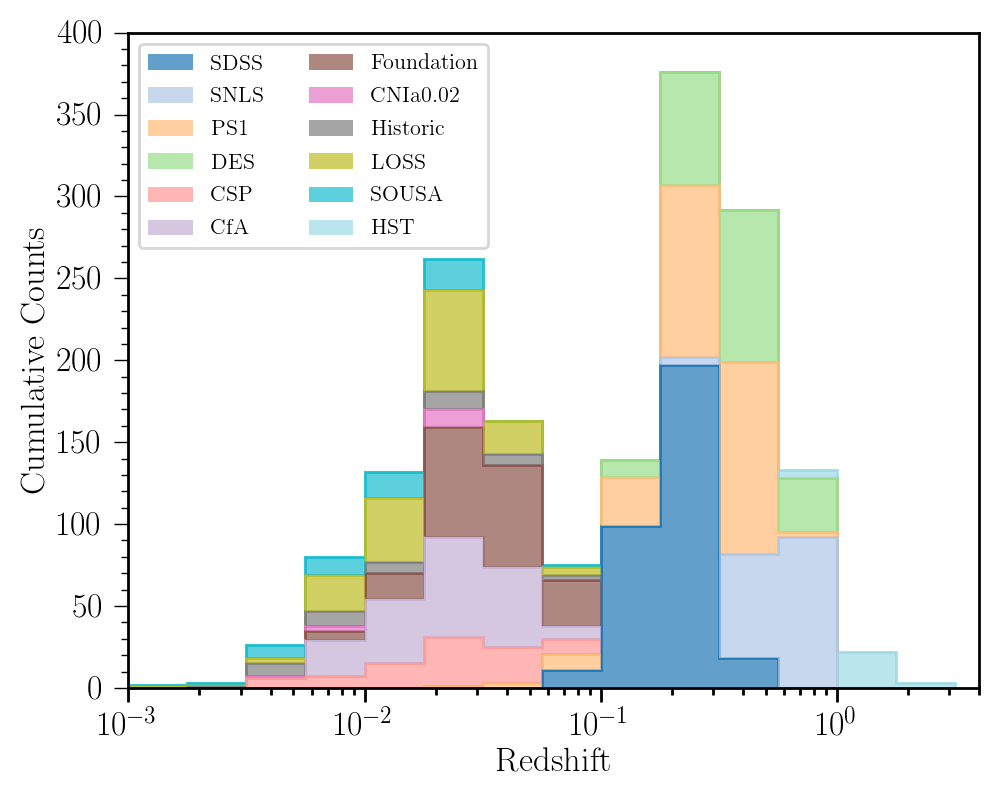}%

  \caption{(Top:) The redshift distribution of the Pantheon+ sample that passes all the light-curve requirements, as well as the same for the JLA and Pantheon samples.  The largest increase in the number of SNe for the Pantheon+ sample is at low redshift owing to the addition of the Foundation, LOSS1, LOSS2, SOUSA, and CNIa0.2 samples.  The largest increase at higher redshift is due to the inclusion of the DES 3-year sample. We do not use SNe from SNLS at $z>0.8$ due to sensitivity to the $U$-band in model training, so the Pantheon+ statistics between $0.8<z<1.0$ are lower than that of Pantheon and JLA. (Bottom:) The Pantheon+ redshift diagram shown cumulatively by survey.}
      
    \label{fig:redshifts} \vspace{-1mm}
\end{figure} 

The Pantheon+ sample comprises 18 different samples, where a sample is loosely defined as the dataset produced by a single SN survey over a discrete period of time.  The samples and their references, as well as their redshift ranges, are given in Table~\ref{tab:surveys}.  In the Appendix, we give an overview of each sample where we detail the original data-release paper, the location of the data, and the photometric system of the SNe.  This table should be combined with the tables in Appendix A in B22a that have the information for the photometric systems and information on stellar catalogs used for cross-calibration.

Here we review the main changes since the first Pantheon release.  We have added 6 large samples: Foundation Supernova Survey \citep[Foundation;][]{Foley2018}, the {\it Swift} Optical/Ultraviolet Supernova Archive (SOUSA)\footnote{Light curves from SOUSA can be found at \url{https://pbrown801.github.io/SOUSA/}.}, the first sample from the Lick Observatory Supernova Search  \citep[LOSS1;][]{Ganeshalingam2010}, the second sample from LOSS \citep[LOSS2;][]{Stahl2019}, and the Dark Energy Survey \cite[DES;][]{Brout19}.  All but DES are low-$z$ surveys, which is why in  Figure~\ref{fig:redshifts} the largest improvement in SN numbers is at low redshift. Additionally, there was a new data release for the Carnegie Supernova Project \citep[CSP;][]{Krisciunas2017} which remeasured previous photometry for CSP-I and added more SNe. 

We note that beyond the main samples included in the data releases described in Table~\ref{tab:surveys}, there are additional light curves here.  These include CSP SNe (SNe~2015F, 2013aa, 2012ht, and 2012fr) from CSP-II that were published by \cite{Burns18} and \cite{Burns20}, SN~2005df from \cite{Milne10} and \cite{Kris2017_df}, SN~2006dd from \cite{Stritzinger10}, SN~2007on and SN~2011iv from \cite{Gall2018}, SN~2007gi from \cite{Zhang10}, SN~2008fv from 
\cite{Tsvetkov10}, SN~2019ein from \cite{Kawabata20}.

Additionally, there are light curves that have not yet been published, but are included in the respective Pantheon+ sample.  These are SN~2021pit from SOUSA and SN~2021hpr from LOSS2, which follow the processing and photometric systems of the larger samples.  Additionally, there are three light curves from Foundation after their release (SN~2017erp, SN~2018gv, and SN~2019np).  SN~2018gv and SN~2019n were processed with the same pipeline described in \cite{Foley2018}.  For SN~2017erp, this SN was outside of the PS1 footprint, so Skymapper catalogs \citep{Onken19} were used to set the photometric zeropoints following the process outlined in \cite{Scolnic2015}.

We have made a special effort to calibrate and include surveys that contain observations of  SNe~Ia in near enough galaxies ($\lesssim40$ Mpc) for which Cepheid observations with the {\it Hubble Space Telescope (HST)} have been obtained because such objects are rare (approximately one per year) and their numbers limit the precision of the determination of H$_0$ (see R22).  As shown by \cite{Brownsberger21}, the sensitivity of measurements of H$_0$ to the photometric calibration of SN light curves depends on whether the relative number of second-rung SNe observed by a survey is similar to the relative number of third-rung SNe observed by that survey.  \cite{Brownsberger21} demonstrate that our current compilation has sufficiently similar numbers so that the impact of potential cross-survey systematics from calibration is $<0.2\%$ in H$_0$.

For each of the samples, the photometric systems are recalibrated by B22a.  Two surveys previously in Pantheon have changed in response to an improved understanding of their photometry. (1) For SDSS, the reported photometry was thought in Pantheon to be in the AB system but was actually in the natural system, so offsets to the photometry of [$-0.06$, 0.02, 0.01, 0.01, 0.01~mag] in $ugriz$ were not applied in Pantheon (the $u$-band usage in SALT2 is minimal, as most SNe discovered by SDSS are at $z>0.1$, outside the usable redshift range for the $u$ band filter). (2) For CfA3K and CfA3S, the photometry of the SNe was assumed in Pantheon to be in the natural system but was actually in the standard system --- this changes the $B$ band by $\sim+0.01$~mag fainter relative to the other bands.

We release the light curves with the photometry as given by the original sources here (though all put in a standard syntax): \url{ https://pantheonplussh0es.github.io/}. The calibration of the samples and derived offsets to the photometric zeropoints given in B22a will be included at the same github page.  Furthermore, we include files to quickly apply calibration definitions and offsets (e.g., the CALSPEC zeropoints needed to define the photometric systems) to fit the light curves.

\begin{deluxetable*}{p{0.3\linewidth}p{1.2cm}p{0.7cm}p{1.2cm}p{0.3\linewidth}}
\tabletypesize{\footnotesize}
\tablecaption{Characteristics of Datasets and List of Improvements} 
\tablehead{\colhead{Source} & \colhead{Abbrev.~~~} & \colhead{$N_{\mathrm{SN}}~$/~$N_{\mathrm{Tot}}~~~$}  & \colhead{$z$ range} &  \colhead{Reference} }
\startdata
Lick Observatory Supernova Search (1998--2008) & LOSS1 & \numKAITs/165 & 0.0020--0.0948 & \citet{Ganeshalingam2010} \\
Lick Observatory Supernova Search (2005--2018) & LOSS2  & \numKAITm/78 & 0.0008--0.082 & \citet{Stahl2019} \\
{\it Swift} Optical/Ultraviolet Supernova Archive & SOUSA & \numSWIFT/120 & 0.0008--0.0616 & \cite{sousa14},~\url{https://pbrown801.github.io/SOUSA/} \\
Carnegie Supernova Project (DR3) & CSP & \numCSP/134 & 0.0035--0.0836 & \citet{Krisciunas2017} \\
Center for Astrophysics (1) & CfA1  & \numCFAa/22 & 0.0032--0.084\phn{} & \citet{Riess99} \\
Center for Astrophysics (2) & CfA2  & \numCFAb/44 & 0.0032--0.084\phn{} &  \citet{jha06} \\
Center for Astrophysics (3) (4Shooter, Kepler-cam) & CfA3S + CfA3K  & \numCFAt/185 & 0.0032--0.084\phn{} & \citet{Hicken2009} \\
Center for Astrophysics (4p1, 4p2) & CfA4  & \numCFAf/94 & 0.0067--0.0745 & \citet{Hicken2012} \\
Low-redshift (various sources) & LOWZ  & \numLOWZ/81 & 0.0014--0.123\phn{} &  
\citet{jha2007}; \citet{Milne10}; \citet{Tsvetkov10}; \cite{Zhang10}; \cite{Contreras2010}; \cite{Kris2017_df}; \citet{Stritzinger2011}; \citet{Wee2018}; \cite{Kawabata20}\\
Complete Nearby (Redshift $<$ 0.02) Sample & CNIa0.02 & \numASASSN/17 & 0.0041--0.0302 & \cite{Chen20} \\
Foundation Supernova Survey & Foundation  & \numFOUND/228 & 0.01--0.1 & \citet{Foley2018} \\
Sloan Digital Sky Survey & SDSS& \numSDSS/499 & 0.0130--0.5540  & \citet{Sako2018} \\
The Panoramic Survey Telescope \& Rapid Response System Medium Deep Survey & PS1MD  & \numPS/370 & 0.0252--0.670\phn{} & \citet{Scolnic2018} \\
SuperNova Legacy Survey & SNLS & \numSNLS/239 & 0.1245--1.06\phn{}\phn{} & \cite{Betoule2014}\\
Dark Energy Survey (3YR) & DES  & \numDES/251 & 0.0176--0.850\phn{} &  \citet{Brout19}; \citet{Smith2020} \\
Hubble Deep Field North (using \textit{HST}) & HDFN & 0/1 & 1.755 & \citet{Gilliland1998}; \citet{Riess2001} \\
Supernova Cosmology Project (using \textit{HST}) & SCP  & \numSCP/8 & 1.014--1.415 & \citet{Suzuki2012} \\
Cosmic Assembly Near Infra-Red Deep Extragalactic Legacy Survey and Cluster Lensing And Supernova survey with Hubble (using \textit{HST}) & CANDELS +CLASH  & \numCANDELS/13 & 1.03--2.26 & \citet{Riess2018}  \\
Great Observatories Origins Deep Survey and Probing Acceleration Now with Supernova (using \textit{HST}) & GOODS +PANS & \numHST/29 & 0.460--1.390  & \citet{Riess2004}; \citet{Riess2007} \\
\enddata

\tablecomments{The different samples included in the Pantheon+ compilation, the number of SNe that are in the cosmology sample and the number from the full sample, the redshift range, and the reference. We provide fitted light-curve parameters for all the light curves with a converged SALT2 fit as part of the data release, but the cosmological analysis is done only with the SNe that pass all the cuts listed in Table 2.}
  \label{tab:surveys}
\end{deluxetable*}


\begin{figure*}[h]
    \centering 
	\includegraphics[page=1,width=.15\textwidth,trim={0 0 8cm 8cm},clip]{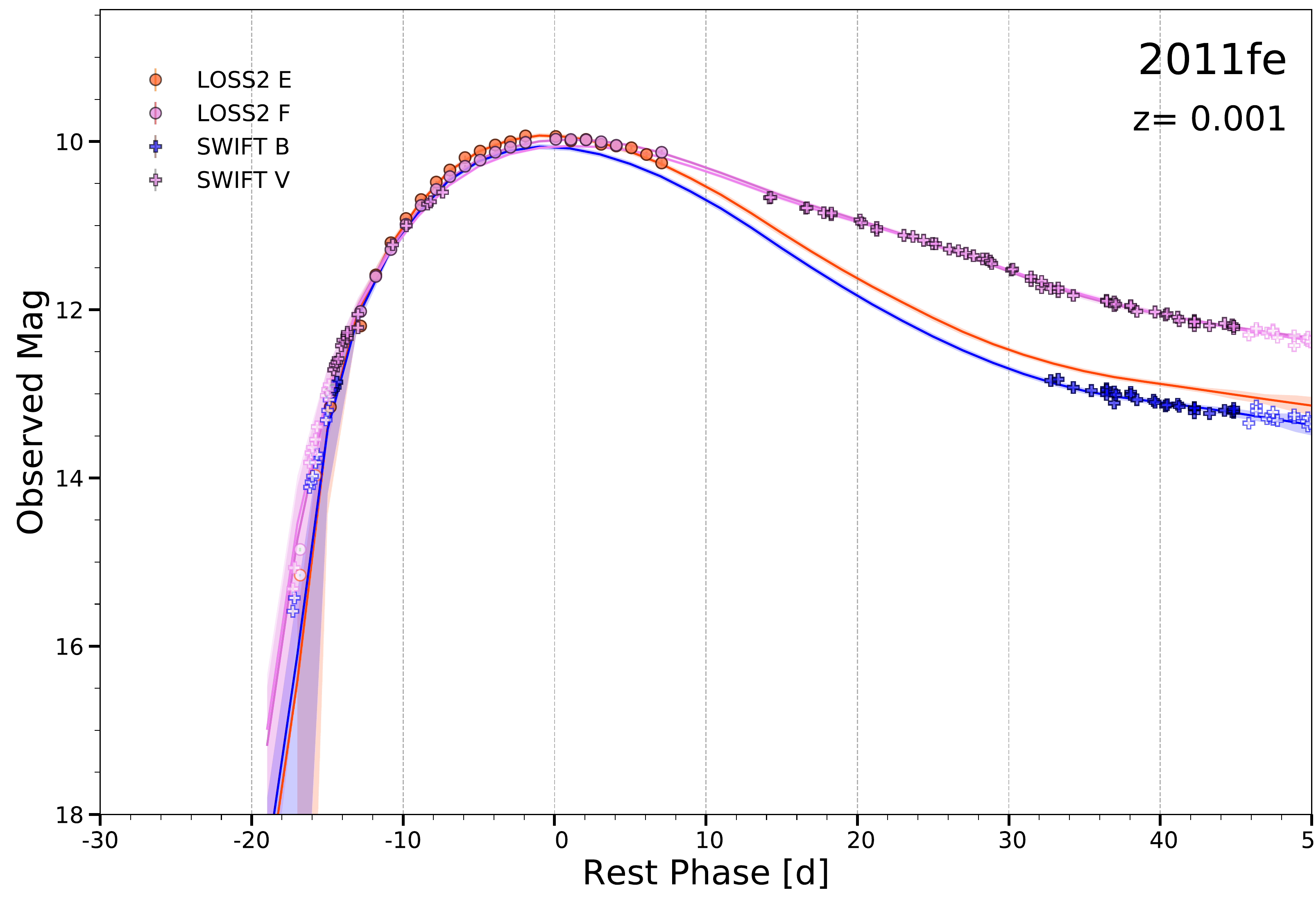} 
	\includegraphics[page=2,width=.15\textwidth,trim={0 0 8cm 8cm},clip]{figures/Calibrator_LightCurves.pdf} 
	\includegraphics[page=3,width=.15\textwidth,trim={0 0 8cm 8cm},clip]{figures/Calibrator_LightCurves.pdf} 
	\includegraphics[page=4,width=.15\textwidth,trim={0 0 8cm 8cm},clip]{figures/Calibrator_LightCurves.pdf} 
	\includegraphics[page=5,width=.15\textwidth,trim={0 0 8cm 8cm},clip]{figures/Calibrator_LightCurves.pdf} 
	\includegraphics[page=6,width=.15\textwidth,trim={0 0 8cm 8cm},clip]{figures/Calibrator_LightCurves.pdf} 
	\includegraphics[page=7,width=.15\textwidth,trim={0 0 8cm 8cm},clip]{figures/Calibrator_LightCurves.pdf} 
	\includegraphics[page=8,width=.15\textwidth,trim={0 0 8cm 8cm},clip]{figures/Calibrator_LightCurves.pdf} 
	\includegraphics[page=9,width=.15\textwidth,trim={0 0 8cm 8cm},clip]{figures/Calibrator_LightCurves.pdf} 
	\includegraphics[page=10,width=.15\textwidth,trim={0 0 8cm 8cm},clip]{figures/Calibrator_LightCurves.pdf} 
	\includegraphics[page=11,width=.15\textwidth,trim={0 0 8cm 8cm},clip]{figures/Calibrator_LightCurves.pdf} 
	\includegraphics[page=12,width=.15\textwidth,trim={0 0 8cm 8cm},clip]{figures/Calibrator_LightCurves.pdf} 
	\includegraphics[page=13,width=.15\textwidth,trim={0 0 8cm 8cm},clip]{figures/Calibrator_LightCurves.pdf} 
	\includegraphics[page=14,width=.15\textwidth,trim={0 0 8cm 8cm},clip]{figures/Calibrator_LightCurves.pdf} 
	\includegraphics[page=15,width=.15\textwidth,trim={0 0 8cm 8cm},clip]{figures/Calibrator_LightCurves.pdf} 
	\includegraphics[page=16,width=.15\textwidth,trim={0 0 8cm 8cm},clip]{figures/Calibrator_LightCurves.pdf} 
	\includegraphics[page=17,width=.15\textwidth,trim={0 0 8cm 8cm},clip]{figures/Calibrator_LightCurves.pdf} 
	\includegraphics[page=18,width=.15\textwidth,trim={0 0 8cm 8cm},clip]{figures/Calibrator_LightCurves.pdf} 
	\includegraphics[page=19,width=.15\textwidth,trim={0 0 8cm 8cm},clip]{figures/Calibrator_LightCurves.pdf} 
	\includegraphics[page=20,width=.15\textwidth,trim={0 0 8cm 8cm},clip]{figures/Calibrator_LightCurves.pdf} 
	\includegraphics[page=21,width=.15\textwidth,trim={0 0 8cm 8cm},clip]{figures/Calibrator_LightCurves.pdf} 
	\includegraphics[page=22,width=.15\textwidth,trim={0 0 8cm 8cm},clip]{figures/Calibrator_LightCurves.pdf} 
	\includegraphics[page=23,width=.15\textwidth,trim={0 0 8cm 8cm},clip]{figures/Calibrator_LightCurves.pdf} 
	\includegraphics[page=24,width=.15\textwidth,trim={0 0 8cm 8cm},clip]{figures/Calibrator_LightCurves.pdf} 
	\includegraphics[page=25,width=.15\textwidth,trim={0 0 8cm 8cm},clip]{figures/Calibrator_LightCurves.pdf} 
	\includegraphics[page=26,width=.15\textwidth,trim={0 0 8cm 8cm},clip]{figures/Calibrator_LightCurves.pdf} 
	\includegraphics[page=27,width=.15\textwidth,trim={0 0 8cm 8cm},clip]{figures/Calibrator_LightCurves.pdf} 
	\includegraphics[page=28,width=.15\textwidth,trim={0 0 8cm 8cm},clip]{figures/Calibrator_LightCurves.pdf} 
	\includegraphics[page=29,width=.15\textwidth,trim={0 0 8cm 8cm},clip]{figures/Calibrator_LightCurves.pdf} 
	\includegraphics[page=30,width=.15\textwidth,trim={0 0 8cm 8cm},clip]{figures/Calibrator_LightCurves.pdf} 
	\includegraphics[page=31,width=.15\textwidth,trim={0 0 8cm 8cm},clip]{figures/Calibrator_LightCurves.pdf} 
	\includegraphics[page=32,width=.15\textwidth,trim={0 0 8cm 8cm},clip]{figures/Calibrator_LightCurves.pdf} 
	\includegraphics[page=33,width=.15\textwidth,trim={0 0 8cm 8cm},clip]{figures/Calibrator_LightCurves.pdf} 
	\includegraphics[page=34,width=.15\textwidth,trim={0 0 8cm 8cm},clip]{figures/Calibrator_LightCurves.pdf} 
	\includegraphics[page=35,width=.15\textwidth,trim={0 0 8cm 8cm},clip]{figures/Calibrator_LightCurves.pdf} 
	\includegraphics[page=36,width=.15\textwidth,trim={0 0 8cm 8cm},clip]{figures/Calibrator_LightCurves.pdf} 
	\includegraphics[page=37,width=.15\textwidth,trim={0 0 8cm 8cm},clip]{figures/Calibrator_LightCurves.pdf} 
	\includegraphics[page=38,width=.15\textwidth,trim={0 0 8cm 8cm},clip]{figures/Calibrator_LightCurves.pdf} 
	\includegraphics[page=39,width=.15\textwidth,trim={0 0 8cm 8cm},clip]{figures/Calibrator_LightCurves.pdf} 
	\includegraphics[page=40,width=.15\textwidth,trim={0 0 8cm 8cm},clip]{figures/Calibrator_LightCurves.pdf} 
	\includegraphics[page=41,width=.15\textwidth,trim={0 0 8cm 8cm},clip]{figures/Calibrator_LightCurves.pdf} 
	\includegraphics[page=42,width=.15\textwidth,trim={0 0 8cm 8cm},clip]{figures/Calibrator_LightCurves.pdf} 
     \caption{Light curves of all SNe~Ia used for the SN~Ia--Cepheid calibration (second rung of the distance ladder).  When a SN has been observed by multiple surveys, multiple light curves are shown for each filter.  The SALT2 fit from each light curve is overplotted. Certain filters (e.g., $I$ and sometimes $R$) are not included in the fit when the observed-frame filter is outside the used SALT2 wavelength range of 300--700~nm. }
    \label{fig:calibrator_lightcurves} \vspace{-1mm}
\end{figure*}

\subsection{Light-Curve Fits}
\label{sec:light curves}

In order to obtain distance moduli ($\mu$) from SN~Ia light curves, we fit the light curves with the SALT2 model (\citealt{Guy07}) using the trained model parameters from B22a over a spectral energy distribution (SED) wavelength range of 200--900~nm. We select passbands whose central wavelength ($\bar\lambda$) 
satisfies 300~nm~$< \bar\lambda/(1+z) < $700~nm, and we select epochs between $-15$ to $+45$ rest-frame days with respect to the epoch
of peak brightness. We use the \texttt{SNANA} software package (\citealt{Kessler2009}) to fit the SALT2 model to the data, and we use SNANA's \texttt{MINOS} computational algorithm to determine the parameters and their uncertainties. 

Each light-curve fit determines the parameters color ($c$), stretch ($x_1$), and overall amplitude ($x_0$), with $m_B \equiv -2.5\log_{10}(x_0)$, as well as the time of peak brightness ($t_0$) in the rest-frame $B$-band wavelength range.  To convert the light-curve fit parameters into a distance modulus, we follow the modified \cite{Tripp} relation as given by \cite{Brout2019}:

\begin{equation}
    \mu=m_B+\alpha x_1 - \beta c - M - \delta_{\mu - \textrm {bias}},
    \label{eqn:Tripp}
\end{equation}
where $\alpha$ and $\beta$ are correlation coefficients, $M$ is the fiducial absolute magnitude of a SN~Ia for our specific standardization algorithm, and $\delta_{\mu - \textrm {bias}}$ is the bias correction derived from simulations needed to account for selection effects and other issues in distance recovery.  For the nominal analysis of B22b, the canonical ``mass-step correction'' $\delta_{\mu - \textrm {host}}$ is  included in the bias correction $\delta_{\mu - \textrm {bias}}$ following \cite{bs20} and \cite{Popovic21}.  The $\alpha$ and $\beta$ used for the nominal fit are 0.148 and 3.112, respectively, and the full set of distance modulus values and uncertainties are presented by B22b.

In addition, we compute a light-curve fit probability ($P_{\rm fit}$), which is the probability of finding a light-curve data-model $\chi^2$ as large or larger assuming Gaussian-distributed flux uncertainties. In Figure~\ref{fig:calibrator_lightcurves}, the light curves of the 42 SNe~Ia used for the determination of H$_0$ in the second-rung distance ladder of R22 are shown with overlaid light-curve fits using the SALT2 model.  All light-curve fit parameters for the sample will be made available in machine-readable format as described in Appendix~\ref{sec:datainfo} and shown in Fig.~\ref{fig:input_fit}.  The parameters from the fits of the light curves are given before the set of light curves before the majority of the selection cuts in Table 2 are applied, which are discussed in the following section.

Finally, in the discussion about the results on siblings and duplicates below, we refer to the distance-covariance matrix.  For this, we follow \cite{Conley2010}, which defines a covariance matrix $C$ with

\begin{equation}
C_{z_iz_j} = \sum_{k} \frac{\partial \Delta\mu_{{\zindex}_i}}{\partial k} ~ \frac{\partial \Delta\mu_{{\zindex}_j}}{\partial k} ~ \sigma_k^2,
\label{eq:cov}
\end{equation}
where the summation is over the systematics ($k$), $\Delta\mu_{{\zindex}_i}$ are the residuals in distance for the SNe fitted between different systematics, and $\sigma_k$ is the gives the magnitude of the systematic uncertainty. Any additional covariance between the $i$th and $j$th SNe that is not due to systematics can be included in that element of the covariance matrix.

\subsection{Selection Requirements}
\label{sec:selection}



For this compilation, we require all SNe~Ia to have adequate light-curve coverage in order to reliably constrain light-curve-fit parameters.  We also limit ourselves to include SNe~Ia with properties in a range well represented by the training sample in order to limit systematic biases in the measured distance modulus. The sequential loss of SNe~Ia from the sample owing to cuts is shown in Table \ref{tab:cuts}.  We define $T_{\rm rest}$ as the number of days since the date of peak brightness $t_0$ in the rest frame of the SN. Following \cite{Scolnic2018}, we require an observation before 5 days after peak brightness ($T_{\rm rest}<5$).  As with \cite{Betoule2014}, we also require the uncertainty in the fitted peak-date of the light-curve (PKMJD)  to be $<2$ observed frame days to ensure precision in the fit. We require $-3 < x_1 < 3$ and $-0.3 < c < 0.3$ over which the light-curve model has been trained. Furthermore, we require that the uncertainty in $x_1$ is $<1.5$ to help avoid pathological fits or inversion issues for systematic uncertainty covariance matrices.

For all samples (though only applicable at low $z$) we require limited Milky Way extinction following \cite{Betoule2014} and \cite{Scolnic2015}, $E(B-V)_{\rm MW}< 0.2$.  We follow past analyses of specific samples in order to employ a minimum $P_{\rm fit}$ cut: this is done for DES, PS1, and SDSS with levels of 0.01, 0.001, and 0.001, respectively.  These different levels are determined from comparisons of distributions of $P_{\rm fit}$ from data and simulations, and depend on the accuracy of the SALT2 model and of the precision of the photometric errors given for SN light-curve measurements.  SNLS is the only large, high-$z$ sample in which a $P_{\rm fit}$ cut is not applied, and this is because \cite{Betoule2014} found no difference in the accuracy of the fitted light curves with low $P_{\rm fit}$.  We see similar insignificant differences in Hubble residuals or fit parameters between SNe with high and low $P_{\rm fit}$ as \cite{Betoule2014} do, but retain the usage of $P_{\rm fit}$ to be consistent with how SNLS was previously used.  Finally, we remove all SNLS and DES SNe from the sample for $z>0.8$, as B22a find large ($\sim0.2$) differences in $\mu$ for these SNe depending on the inclusion of the $U$ band at low redshift in the SALT2 training samples, and we are unable to calibrate $U$ through cross-calibration.  In total, 59 SNe are removed owing to this cut.

\begin{deluxetable}{p{3.5cm} l p{1cm}}
\tablecaption{Cosmology sample cuts}
\tablehead{\colhead{Cut} & \colhead{Discarded} & \colhead{Remaining}}
\startdata
SALT2 converged & - & 2077 \\
$P_{\rm fit}$ & 16 & 2061 \\
$U$-band sensitivity & 59 & 2002 \\
$\sigma(x_1) < 1.5$ & 85 & 1917 \\
$\sigma_{\rm (pkmjd)} < 2 $ & 10 & 1907 \\
$-0.3 < c < 0.3$ & 98 & 1909 \\
$-3 < x_1 < 3$ & 7 & 1802 \\
$ E(B-V)_{\text{MW}}< 0.20$ mag & 23 & 1779 \\
$T_{\text{rest}}<5$ & 1 & 1778 \\
\hline
Chauvenet's criterion & 5 & 1773 \\
Valid BiasCor & 10 & 1763 \\
Systematics  & 60 & 1701 \\
\enddata
\tablecomments{Impact of various cuts used for cosmology analysis.  Both the number removed from each cut, and the number remaining after each cut, are shown.  The ``SALT2 converged" criterion is the starting point for this assessment and includes all light curves for which the fitting procedure converged.  Of the 1701 light curves that pass all cuts, 151 are ``Duplicate'' SNe.}
\label{tab:cuts}
\end{deluxetable}

In the penultimate row of Table~\ref{tab:cuts} (``Valid BiasCor''), 10 light curves are lost owing to their light-curve properties falling within a region of parameter space that is too sparsely populated in the simulation to yield a meaningful bias prediction. Bias corrections are discussed in detail by B22b. Additionally, there are 60 more light curves that are lost owing to the requirement that they pass all the cuts discussed above for the 40 systematic perturbations discussed by B22b in order to create the covariance matrix in Equation~2. For example, varying the SALT2 model will change the recovered $c$ or $x_1$ values, which could then be outside the allowed ranges. Additionally, B22b place a cut on SN distance modulus values in the Hubble diagram due to Chauvenet's criterion.  We label the number cut here in Table~\ref{tab:cuts}, and this is discussed in detail by B22b.

In total, 1701 light curves pass all the cuts, though as discussed below, a significant fraction of these are duplicate SNe.

\subsection{Host-galaxy Properties}
\label{hostmass}

In order to allow the use of host-galaxy information that may improve light-curve standardization (e.g., \citealp{Sullivan10,Kelly10,lampeitl10,Popovic21}), we rederived host properties for all SNe~Ia with $z<0.15$ so that they can be measured consistently.  For $z>0.15$ and higher-$z$ surveys, we use the masses provided from respective analyses: for SNLS \citet{Betoule2014}, for SDSS \citet{Sako2018}, for PS1 \citet{Scolnic2018}, and for DES \citet{Smith2020}.  We discuss consistency across these different samples below.  For the \textit{HST} surveys as listed in Table~\ref{tab:surveys}, masses were not originally derived for the majority of the host galaxies, so we followed a similar procedure as below but using photometry directly from the publicly available images acquired as part of the surveys given in Table~\ref{tab:surveys}.

There are three steps we follow to  determine the masses of the host galaxies:

\begin{enumerate}
\item Identify the host galaxy.
\item Measure photometry of the host galaxy.
\item Fit a galaxy SED model to the data.
\end{enumerate}

\noindent For the low-$z$ sample, for host-galaxy identification, we followed the work of C22 to identify host galaxies and used the directional-light-radius method described by \cite{Sullivan06} and \cite{Gupta16} to associate a host galaxy with each SN~Ia.  All host-galaxy identifications were visually inspected for quality control.  We then retrieved images from {\it GALEX} \citep{Galex05}, PS1 \citep{Chambers17}, SDSS \citep{Ahumada20}, and 2MASS \citep{2mass06}. We measure aperture photometry on the images, and use the PS1 $r$ band to measure the size of the host galaxy ``ellipse.''  We then use that ellipse size to measure consistent elliptical aperture photometry for every image of the source. We use $ugriz$ SDSS photometry rather than $griz$ PS1 photometry when both are available as PS1 has some background-subtraction defects for bright hosts \citep{Jones19}.

   
In order to determine host-galaxy properties from the photometry of the galaxies, we used the LePHARE SED-fitting method \citep{lephare06}. The galaxy templates use the \cite{Chabrier03} initial mass function and were taken from the \cite{Bruzal03} library. The values of the extinction $E(B-V$) varied from 0 to 0.4~mag. For galaxies that \texttt{LePHARE} was not able to determine a host mass, we first confirm that the hosts are faint and have not been misidentified, and then we assign them to the low-mass bin. 

\begin{figure}[h] 
\centering 
	\includegraphics[width=.51\textwidth]{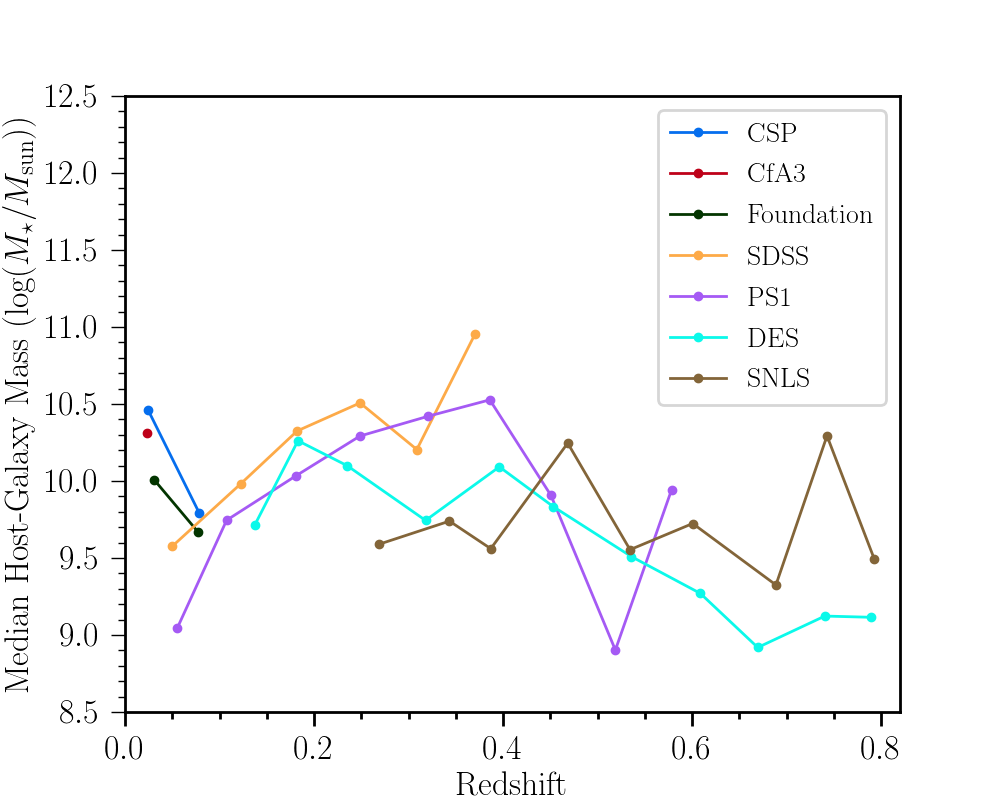} 
   \caption{The median host-galaxy mass per redshift bin for the 7 samples with the highest statistics.}
   \label{fig:submass} \vspace{-1mm}
\end{figure}

A plot of the trend of host-galaxy masses for our largest samples (CSP, Foundation, CfA3, DES, SDSS, SNLS, PS1) is shown in Figure~\ref{fig:submass}.  When we compare different estimates of host-galaxy mass from varying the photometry or mass-fitting technique, we find typical differences on the level of $0.2$~dex (see, e.g., \citealp{Sako2018,Smith2020}), which would make up some of the differences between median mass of different samples. Another way to quantify this is to measure the difference in the relative ratio of high-mass to low-mass hosts (where the separator is $10^{10}$~M$_\odot$) between different surveys.  Doing so, we find that typical differences in the same bin between surveys on the order of $15\%$ would cause $\sim0.01$ mag biases for a mass step of $0.06$ mag if they were systematic and not random.  As there is no evidence of systematic biases beyond the $0.2$~dex scale, this number is used to account for systematics in B22b.  Furthermore, we find relatively good agreement with past estimates compiled in Pantheon, with the typical differences between median masses in the same bin on the level of $0.5$~dex.

\subsection{Trends of SN Parameters and Comparison to Previous Analyses}

\begin{figure}[h]
    \centering 
	\includegraphics[width=.52\textwidth]{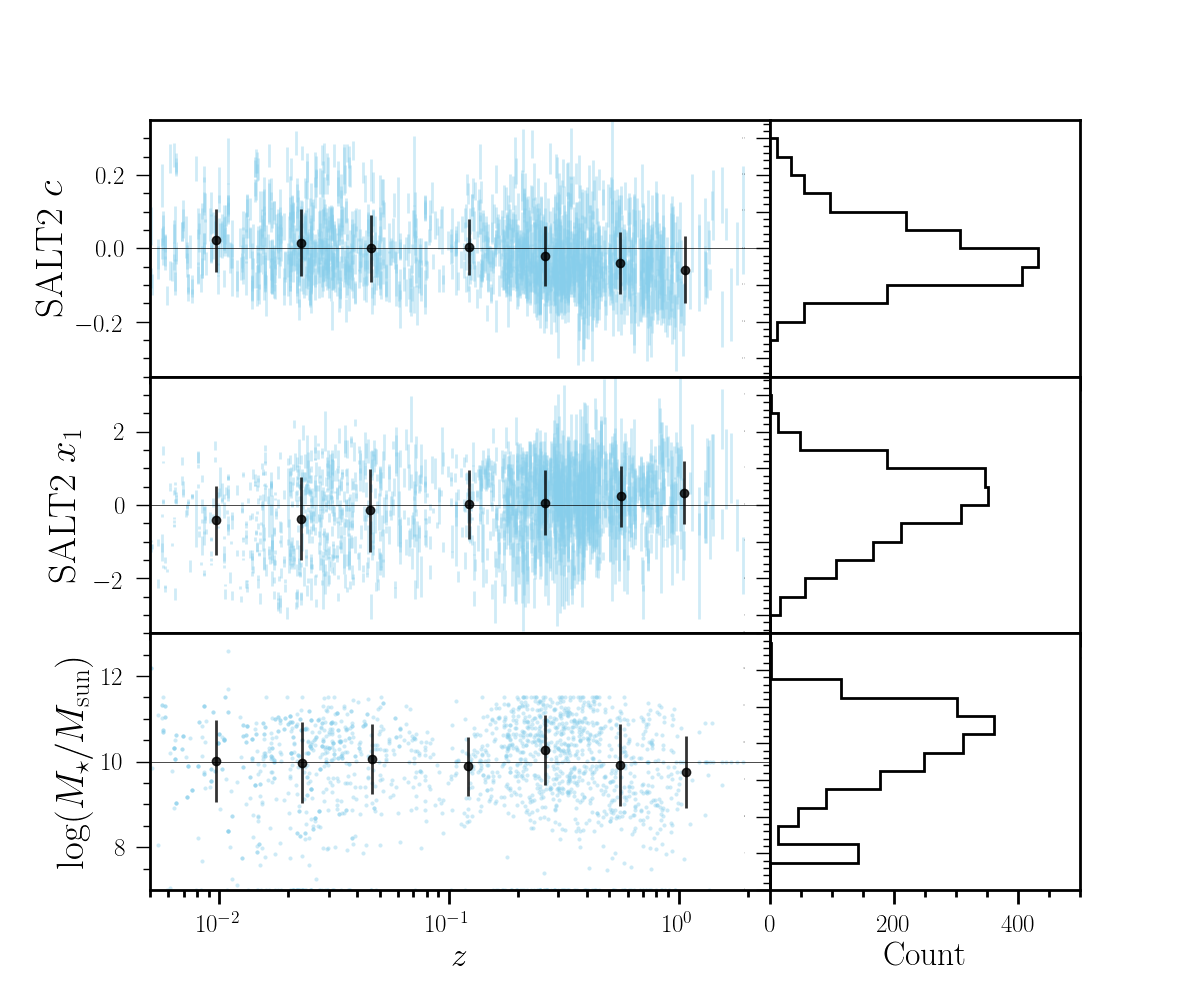} 
      \caption{The evolution of fitted SALT2 $x_1$ and $c$ parameters, as well as the host-galaxy mass with redshift.  SNe are shown here if they pass the light-curve selection requirements discussed in Sec.~\ref{sec:selection}.  The error bars shown here are given as the root-mean square of values in that redshift bin.}
    \label{fig:evol} \vspace{-1mm}
\end{figure}

We show the evolution of the light-curve fit parameters with redshift in Figure~\ref{fig:evol}.  As seen in previous analyses, we do find nonzero evolution of these parameters with redshift.  These are modeled by \cite{Popovic21}, who describe a separate mass distribution for low-$z$  (e.g., CfA1--4, CSP) and high-$z$ (SDSS, SNLS, PS1, DES) samples.

In total, there are \totsn~SNe, significantly more than the number from Pantheon (1048) or JLA (742).  All but 14 of the SNe in Pantheon are in Pantheon+, and all but 10 of the SNe in JLA are in Pantheon+. In B22a, we show the differences between the $\mu$ values found in Pantheon+ and those found in Pantheon and JLA. The largest differences are due to the calibration of the SALT2 model, which is revised by B22a.
We note that the issue of revising the CfA3K and CfA3S system definition mentioned previously does cause a $\sim0.025$ mag change (toward fainter distance-modulus values) relative to Pantheon.

   


\section{Sibling and Duplicate Supernovae }

\begin{deluxetable*}{lrrrrrr}
\tablecaption{Table of Supernova Siblings}
\tablehead{\colhead{Supernova} & \colhead{$m_B\pm \sigma_{mB}$} & \colhead{$c\pm \sigma_c$} & \colhead{$x_1\pm \sigma_{x1}$} & \colhead{$\mu -M \pm \sigma_{\mu}$} & \colhead{$z$CMB} & \colhead{Sample}}
\startdata
2010gp       & 15.89$\pm$0.06  & 0.17$\pm$0.05   & -0.51$\pm$0.25  & 15.21$\pm$0.14  & 0.024 & SOUSA \\
PS1-14xw     & 15.57$\pm$0.06  & -0.04$\pm$0.06  & 1.23$\pm$0.71   & 15.86$\pm$0.18  & 0.024 & SOUSA \\
\hline
2007sw       & 15.98$\pm$0.04  & 0.09$\pm$0.03   & 0.08$\pm$0.13   & 15.67$\pm$0.07  & 0.025 & CfA4p1 \\
2012bh       & 15.92$\pm$0.09  & -0.04$\pm$0.05  & -0.41$\pm$0.26  & 15.96$\pm$0.12  & 0.025 & LOSS2 \\
\hline
2007on       & 12.70$\pm$0.02   & 0.00$\pm$0.02    & -2.22$\pm$0.04  & 12.41$\pm$0.06  & 0.006 & SOUSA\\
2011iv       & 12.11$\pm$0.02  & -0.06$\pm$0.02  & -1.90$\pm$0.04   & 11.97$\pm$0.05  & 0.006 & CSP \\
\hline
2007sw       & 15.98$\pm$0.04  & 0.09$\pm$0.03   & 0.08$\pm$0.13   & 15.67$\pm$0.07  & 0.025 & CfA4p1\\
370356       & 15.81$\pm$0.05  & -0.08$\pm$0.04  & -0.29$\pm$0.13  & 15.99$\pm$0.08  & 0.025 & PS1 \\
\hline
1980N        & 12.09$\pm$0.03  & -0.01$\pm$0.03  & -1.14$\pm$0.12  & 12.00$\pm$0.07   & 0.006 & LOWZ\\
1981D        & 12.22$\pm$0.06  & 0.16$\pm$0.06   & -1.15$\pm$0.36  & 11.66$\pm$0.14  & 0.006 & LOWZ \\
2006dd       & 11.99$\pm$0.03  & 0.02$\pm$0.03   & -0.29$\pm$0.04  & 11.93$\pm$0.06  & 0.006 & LOWZ\\
\hline
2000dk\_v1       & 15.07$\pm$0.05  & -0.04$\pm$0.03  & -2.04$\pm$0.08  & 14.90$\pm$0.06   & 0.016  & LOSS1\\
2000dk\_v2      & 15.00$\pm$0.05   & -0.09$\pm$0.04  & -2.44$\pm$0.1   & 14.89$\pm$0.11  & 0.016 & CfA2 \\
2015ar & 14.74$\pm$0.06  & -0.10$\pm$0.04   & -1.98$\pm$0.19  & 14.71$\pm$0.1   & 0.016 & Foundation \\
\hline
1994M        & 15.98$\pm$0.04  & 0.04$\pm$0.03   & -1.43$\pm$0.09  & 15.71$\pm$0.07  & 0.024 & CfA1\\
2004br       & 15.12$\pm$0.03  & -0.11$\pm$0.03  & 1.00$\pm$0.08    & 15.57$\pm$0.06  & 0.024 & LOSS1\\
\hline
1999cp\_v1       & 13.62$\pm$0.03  & -0.11$\pm$0.03  & 0.29$\pm$0.08   & 13.96$\pm$0.06  & 0.010 & LOSS1 \\
1999cp\_v2      & 13.63$\pm$0.03  & -0.05$\pm$0.04  & 0.02$\pm$0.04   & 13.79$\pm$0.1   & 0.010 & LOWZ \\
2002cr\_v1       & 13.92$\pm$0.03  & -0.04$\pm$0.03  & -0.37$\pm$0.06  & 14.0$\pm$0.06   & 0.010 & LOSS1\\
2002cr\_v2       & 13.86$\pm$0.02  & -0.07$\pm$0.02  & -0.6$\pm$0.03   & 13.99$\pm$0.04  & 0.010 & CfA3S\\
\hline
2013aa\_v1       & 10.81$\pm$0.1   & -0.15$\pm$0.04  & 0.60$\pm$0.1     & 11.29$\pm$0.06  & 0.005 & SOUSA \\
2013aa\_v2       & 10.84$\pm$0.11  & -0.11$\pm$0.05  & 0.51$\pm$0.15   & 11.21$\pm$0.11  & 0.005 & CSP \\
2017cbv\_v1      & 10.86$\pm$0.1   & -0.10$\pm$0.04   & 0.82$\pm$0.07   & 11.27$\pm$0.06  & 0.005 & CSP \\
2017cbv\_v2      & 10.68$\pm$0.09  & -0.14$\pm$0.03  & 0.6$\pm$0.04    & 11.14$\pm$0.05  & 0.005 & CNIa0.02 \\
\hline
2001el       & 12.42$\pm$0.03  & 0.07$\pm$0.03   & -0.13$\pm$0.03  & 12.25$\pm$0.06  & 0.004 & LOWZ \\
2021pit      & 12.03$\pm$0.04  & 0.07$\pm$0.04   & -0.04$\pm$0.12  & 11.75$\pm$0.1   & 0.004 & SOUSA\\
\hline
2013fa       & 15.32$\pm$0.06  & 0.20$\pm$0.03    & -0.56$\pm$0.09  & 14.54$\pm$0.06  & 0.014 & LOSS2\\
PSN J20435314$+$1230304  & 15.68$\pm$0.07  & 0.09$\pm$0.04   & -2.55$\pm$0.15  & 14.94$\pm$0.1   & 0.014 & Foundation\\
\hline
2021hpr      & 13.98$\pm$0.03  & 0.04$\pm$0.03   & 0.25$\pm$0.07   & 13.85$\pm$0.06  & 0.010 & LOSS2\\
1997bq       & 14.10$\pm$0.04   & 0.08$\pm$0.03   & -0.61$\pm$0.09  & 13.82$\pm$0.06  & 0.010  & CfA2 \\
2008fv       & 14.22$\pm$0.03  & 0.11$\pm$0.03   & 0.74$\pm$0.06   & 13.93$\pm$0.06  & 0.010 & LOWZ\\
\enddata
\tablecomments{The Tripp parameters as well as the distance-modulus values $\mu$ minus the absolute magnitude $M$ for each SN sibling as part of a pair or triplet, where each group is separated by a horizontal line in the table. The uncertainties in $\mu$ do not include contributions from peculiar velocities or intrinsic scatter.  Additionally, the sample source of the SN is given in the last column, and we include measurements of the same SN from multiple samples where available.}
\label{table:siblings}
\end{deluxetable*}

\subsection{Sibling Supernovae}

As part of this analysis and that of C22, we have determined the host galaxy for each SN in our sample.  We can then query for galaxies that have hosted more than one SN that make it to the Hubble diagram.  Note that owing to our strict quality cuts, this number is fewer than the total number of SN siblings.  We find 12 galaxies that have hosted SN siblings, as listed in Table~\ref{table:siblings}.  We include the measurements from different samples if a SN has been observed by multiple telescopes.  Two of the galaxies hosted three SNe, and we consider all pair-wise combinations of the triplets.

Comparing the properties of the SNe, we find the standard deviation of the differences in $c$ of 0.10, in $x_1$ of 1.04, and in $\mu$ of 0.32~mag.  We can compare these values to those taking random pairs of SNe at low $z$ by bootstrapping: $c$ of 0.12, $x_1$ of 1.6, and $\Delta \mu$ of 0.22, where $\Delta \mu$ subtracts off the best-fit cosmology to account for two SNe having two different redshifts. A median $0.22$~mag difference is consistent with expectations for SNe with a dispersion of $\sim0.16$~mag, which is the RMS on the Hubble diagram found in B22b. We find that the uncertainties in the standard deviation are $0.023$ in $c$, $0.33$ in $x_1$, and $0.043$ in $\Delta \mu$.  Therefore, we find that the $x_1$ values for the siblings are $\sim2\sigma$ closer than two random SNe, the $c$ values are $<1\sigma$ closer, but the $\mu$ values are $2.4\sigma$ further apart in the siblings than any random pair of SNe. The relatively high agreement in $x_1$ but low agreement in $\Delta \mu$ is consistent with the findings of \cite{Scolnic20} for 8 pairs of siblings found in the DES sample: there are indications that $x_1$ is correlated for SNe in the same hosts, but no significant evidence that the $\Delta \mu$ values are correlated.  This insight is important for creating the systematic covariance matrix of B22b that no covariance should be given for measurements of SN distances in the same galaxy.

\subsection{Duplicate Supernovae}

\begin{figure}
    \centering 
	\includegraphics[width=.49\textwidth]{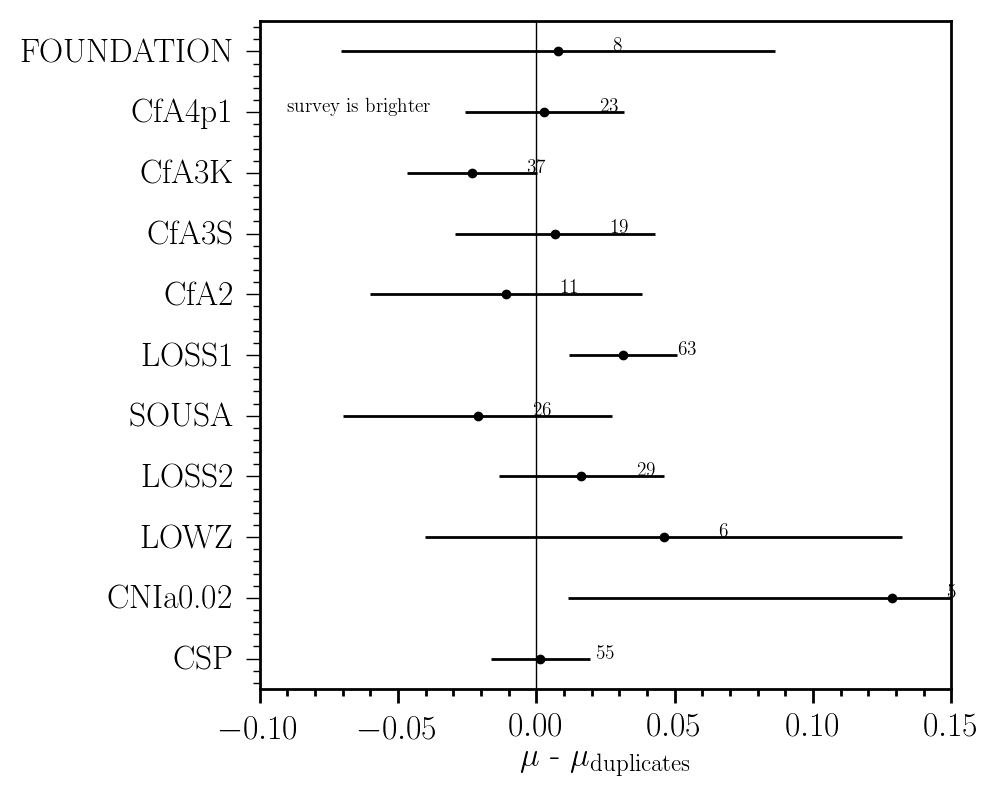} 
      \caption{A comparison of the duplicate SNe.  Here we show the mean difference in distance modulus of duplicate SNe in a given survey and all the other SNe with duplicate SNe observed.  The number of duplicates is labeled in the plot.  No offset is beyond $2\sigma$.}
    \label{fig:duplicates} \vspace{-1mm}
\end{figure} 

We denote SNe that have been observed by multiple surveys as ``Duplicate SNe.''  As discussed by R22 and B22b, unlike in previous analyses, we do not choose between specific versions of the SNe and instead propagate each fit from each survey, and then include a covariance term between the duplicate SNe in our final covariance matrix used for cosmology.  Not all duplicate SNe have the same given name, and we therefore search on RA, DEC, and PKMJD for duplicate SNe.  In total, there are 151 SNe which have been observed by more than one survey, with all but one duplicate SN having $z<0.1$.

We find a standard deviation of the differences in the pairs of $0.102$ mag.  Following a similar bootstrapping procedure as above, and only using low-$z$ SNe, we calculate a typical dispersion for $151$ pairs of random SNe (correcting for redshift differences) to have $0.218$~mag with $0.011$ uncertainty.  Therefore, the distances of the same SN measured by two separate surveys agree by $>10\sigma$ better than two random SNe.  This insight is again important for creating the systematic covariance matrix in B22b that the intrinsic scatter of a SN~Ia should be shared for measurements of the same SN by different surveys; from Equation~2, $C_{{\zindex}_i,{\zindex}_i}=\sigma_{\rm int}$, where the $i$-th and $j$-th light curve are of the same SN from different samples, and $\sigma_{\rm int}$ is the intrinsic scatter of the sample.

In Figure~\ref{fig:duplicates}, we present a comparison of the distance modulus of the SN duplicates between surveys.  We do not find any deviations from the mean beyond $2\sigma$.  The largest deviation is from LOSS1 \citep{Ganeshalingam2010} at $2.0\sigma$.  B22a show the mean distance modulus residuals for each subsample for all surveys and do not find any magnitude deviations greater than $0.05$~mag with the exception of CfA1. Our results here generally support the agreement found by B22a. 

Furthermore, in Table~\ref{table:h0impact}, we give the fraction of the sample each survey contributes to the 2nd and 3rd rungs of the distance ladder described in R22, where the 3rd rung has the limit of $z<0.15$ and those in the 2nd rung are determined SNe found in nearby galaxies with associated Cepheid measurements.  (We note that the baseline determination of H$_0$ further limits the 3rd rung sample to $z>0.0233$ and to late-type hosts.)  Assuming (gray) survey errors, an estimate of the error in H$_0$ from survey miscalibration results from the {\it difference} in these fractions multiplied  by the mean residual of each survey from the full compilation.    We give the fractional difference between these two rungs by sample and the survey residual calculated by B22a (see Figure 6) in Table~\ref{table:h0impact}. If one multiplies the fractional difference between rungs by the Hubble residual offsets, this describes the sensitivity of $H_0$ (in magnitudes, not km/s/Mpc) to possible discrepancies of sample offsets. We find that the largest fractional difference is due to Foundation at $\sim23\%$, and the majority of the fractional differences are between $2-15\%$.  After multiplying these differences by the Hubble residual offsets, we find the products are all below 4 mmag.  This would imply a sensitivity in $H_0$ on the level of $0.2\%$.  This also illustrates the benefit of using a similar mix of surveys for both samples.  Because we cannot avoid using a mix of surveys for the 2nd rung (these are objects are rare) the use of a single sample for the 3rd rung would propagate an error in H$_0$ at the level of $\sim$ 1\% as shown in \citet{Brownsberger21}.

\begin{deluxetable*}{lrrrrr}
\tablecaption{Fraction of SNe in 2nd and 3rd rungs of distance ladder}
\tablehead{\colhead{SURVEY} & \colhead{Frac. HF (rung 3)} & \colhead{Frac. CAL (rung 2)} & \colhead{$\Delta$ (3-2)} & \colhead{$\Delta \mu$} & \colhead{$\Delta \mu \times \Delta$ (3-2) (mag)} }
\startdata
SDSS & 0.112 & 0.000 & -0.112 & 0.009 & -0.0010 \\
PS1 & 0.054 & 0.000 & -0.054 & 0.036 & -0.0019 \\
DES & 0.010 & 0.000 & -0.010 &-0.015 &0.0001 \\
CSP & 0.083 & 0.129 & 0.046 & 0.034 & 0.0015 \\
CfA1 & 0.018 & 0.048 & 0.030 & -0.104 & -0.0031 \\
CfA2 & 0.026 & 0.048 & 0.021 & -0.023 & -0.0005 \\
CfA3 & 0.089 & 0.077 & -0.012 & -0.001 & 0.0000\\
CfA4 & 0.054 & 0.016 & -0.038 & 0.033 & -0.0012 \\
Foundation & 0.279 & 0.052 & -0.227 & 0.008 & -0.0017 \\
CNIa0.02 & 0.018 & 0.040 & 0.022 & 0.027 & 0.0006 \\
LOWZ &  0.056 & 0.187 & 0.131 & -0.020 & -0.0002 \\
LOSS & 0.147 &0.244 & 0.097 & 0.012 & 0.0012 \\
SOUSA & 0.057 & 0.161 & 0.105 & 0.005 & 0.0005 \\
\enddata
\tablecomments{The relative fractions of SN samples by survey (accounting for duplicates) for the 3rd rung of distance ladder ($z<0.15$) SN sample, the 2nd rung of distance ladder Cepheid-hosted SN sample in R21, the difference between the two, the mean offset by survey given in B22a in Fig 6, and the product of the survey offset with the fractional difference.  The product indicates the size of the sensitivity of H$_0$ (in mag, not km/s/Mpc - divide by $\sim2$ for $\%$ units in H$_0$) to survey mis-calibration or other issues. See \cite{Brownsberger21} for more information about this sensitivity.}
\label{table:h0impact}
\end{deluxetable*}

\section{Discussion and Conclusions}

In this paper, we presented the new ``Pantheon+'' sample that is used in a series of analyses for cosmological parameter measurements.  The challenge of a compilation analysis like this one is documentation, and unlike previous analyses, we attempt here to document key properties about the samples (photometric system, data location, references) to improve reproducibility in the future.

The Pantheon+ analysis improves on the Pantheon analysis in nearly every facet.  Not only do we increase the sample size, but we do a comprehensive review of the redshifts (C22) and peculiar velocities \citep{Peterson2021}, a new calibration and model retraining for the sample (B22a), and new cosmological analyses by R22 and B22b.  In Section 2, we detail data that have been added to the previous Pantheon compilation, as well as changes to the data that were previously used.  As these samples date back 40~yr, we have made a significant effort to check assumptions about how data have been passed from analysis to analysis, rather than assuming previous analyses have understood each facet correctly.

The size of a sample like this will soon be surpassed by other samples from newer and upcoming surveys like the Zwicky Transient Facility (ZTF; \citealp{dhawan21}), the Young Supernova Experiment (YSE; \citealp{jones21}), the Dark Energy Survey (DES; \citealp{Smith2020}), the Legacy Survey of Space and Time (LSST; \citealp{zeljko19}), and the Nancy Grace Roman Space Telescope (Roman; \citealp{Hounsell18}). These surveys may find a similar number of SNe to this compilation in only a matter of days.  {\it However}, the usefulness of the Pantheon+ sample, particularly at low redshift, is unlikely to be surpassed for some time owing to its utility for constraining the Hubble constant.  For this measurement, we are statistically limited by the number of SNe in nearby galaxies in which Cepheids can be found, which is typically one SN discovered per year (R22).  

Two of the findings from this paper will be used to create the systematic covariance matrix of B22b.  The first is that we find excellent agreement when different surveys measure the same SNe, and the second is that we find relatively poor agreement when surveys measure distances of two SNe in the same galaxy.  The latter of these findings will be best tested with LSST, which can find over 800 siblings \citep{Mandelbaum18,Scolnic20}.  Finally, we show that because of our effort to include samples that cover the second and third rung of the distance ladder, the accuracy of the $H_0$ measurement will not be limited by possible discrepancies in measurements of the SN distances by sample.

\bibliography{paper}

\begin{thebibliography}{}
\expandafter\ifx\csname natexlab\endcsname\relax\def\natexlab#1{#1}\fi
\providecommand{\url}[1]{\href{#1}{#1}}
\providecommand{\dodoi}[1]{doi:~\href{http://doi.org/#1}{\nolinkurl{#1}}}
\providecommand{\doeprint}[1]{\href{http://ascl.net/#1}{\nolinkurl{http://ascl.net/#1}}}
\providecommand{\doarXiv}[1]{\href{https://arxiv.org/abs/#1}{\nolinkurl{https://arxiv.org/abs/#1}}}

\bibitem[{{Ahumada} {et~al.}(2020){Ahumada}, {Prieto}, {Almeida}, {Anders},
  {Anderson}, {Andrews}, {Anguiano}, {Arcodia}, {Armengaud}, {Aubert}, {Avila},
  {Avila-Reese}, {Badenes}, {Balland}, {Barger}, {Barrera-Ballesteros}, {Basu},
  {Bautista}, {Beaton}, {Beers}, {Benavides}, {Bender}, {Bernardi}, {Bershady},
  {Beutler}, {Bidin}, {Bird}, {Bizyaev}, {Blanc}, {Blanton}, {Boquien},
  {Borissova}, {Bovy}, {Brandt}, {Brinkmann}, {Brownstein}, {Bundy}, {Bureau},
  {Burgasser}, {Burtin}, {Cano-D{\'\i}az}, {Capasso}, {Cappellari}, {Carrera},
  {Chabanier}, {Chaplin}, {Chapman}, {Cherinka}, {Chiappini}, {Doohyun Choi},
  {Chojnowski}, {Chung}, {Clerc}, {Coffey}, {Comerford}, {Comparat}, {da
  Costa}, {Cousinou}, {Covey}, {Crane}, {Cunha}, {Ilha}, {Dai}, {Damsted},
  {Darling}, {Davidson}, {Davies}, {Dawson}, {De}, {de la Macorra}, {De Lee},
  {Queiroz}, {Deconto Machado}, {de la Torre}, {Dell'Agli}, {du Mas des
  Bourboux}, {Diamond-Stanic}, {Dillon}, {Donor}, {Drory}, {Duckworth},
  {Dwelly}, {Ebelke}, {Eftekharzadeh}, {Davis Eigenbrot}, {Elsworth},
  {Eracleous}, {Erfanianfar}, {Escoffier}, {Fan}, {Farr},
  {Fern{\'a}ndez-Trincado}, {Feuillet}, {Finoguenov}, {Fofie},
  {Fraser-McKelvie}, {Frinchaboy}, {Fromenteau}, {Fu}, {Galbany}, {Garcia},
  {Garc{\'\i}a-Hern{\'a}ndez}, {Oehmichen}, {Ge}, {Maia}, {Geisler}, {Gelfand},
  {Goddy}, {Gonzalez-Perez}, {Grabowski}, {Green}, {Grier}, {Guo}, {Guy},
  {Harding}, {Hasselquist}, {Hawken}, {Hayes}, {Hearty}, {Hekker}, {Hogg},
  {Holtzman}, {Horta}, {Hou}, {Hsieh}, {Huber}, {Hunt}, {Chitham}, {Imig},
  {Jaber}, {Angel}, {Johnson}, {Jones}, {J{\"o}nsson}, {Jullo}, {Kim},
  {Kinemuchi}, {Kirkpatrick}, {Kite}, {Klaene}, {Kneib}, {Kollmeier}, {Kong},
  {Kounkel}, {Krishnarao}, {Lacerna}, {Lan}, {Lane}, {Law}, {Le Goff}, {Leung},
  {Lewis}, {Li}, {Lian}, {Lin}, {Long}, {Longa-Pe{\~n}a}, {Lundgren}, {Lyke},
  {Ted Mackereth}, {MacLeod}, {Majewski}, {Manchado}, {Maraston}, {Martini},
  {Masseron}, {Masters}, {Mathur}, {McDermid}, {Merloni}, {Merrifield},
  {M{\'e}sz{\'a}ros}, {Miglio}, {Minniti}, {Minsley}, {Miyaji}, {Mohammad},
  {Mosser}, {Mueller}, {Muna}, {Mu{\~n}oz-Guti{\'e}rrez}, {Myers}, {Nadathur},
  {Nair}, {Nandra}, {do Nascimento}, {Nevin}, {Newman}, {Nidever}, {Nitschelm},
  {Noterdaeme}, {O'Connell}, {Olmstead}, {Oravetz}, {Oravetz}, {Osorio},
  {Pace}, {Padilla}, {Palanque-Delabrouille}, {Palicio}, {Pan}, {Pan},
  {Parker}, {Paviot}, {Peirani}, {Ram{\'r}ez}, {Penny}, {Percival},
  {Perez-Fournon}, {P{\'e}rez-R{\`a}fols}, {Petitjean}, {Pieri},
  {Pinsonneault}, {Poovelil}, {Povick}, {Prakash}, {Price-Whelan}, {Raddick},
  {Raichoor}, {Ray}, {Rembold}, {Rezaie}, {Riffel}, {Riffel}, {Rix}, {Robin},
  {Roman-Lopes}, {Rom{\'a}n-Z{\'u}{\~n}iga}, {Rose}, {Ross}, {Rossi},
  {Rowlands}, {Rubin}, {Salvato}, {S{\'a}nchez}, {S{\'a}nchez-Menguiano},
  {S{\'a}nchez-Gallego}, {Sayres}, {Schaefer}, {Schiavon}, {Schimoia},
  {Schlafly}, {Schlegel}, {Schneider}, {Schultheis}, {Schwope}, {Seo},
  {Serenelli}, {Shafieloo}, {Shamsi}, {Shao}, {Shen}, {Shetrone}, {Shirley},
  {Aguirre}, {Simon}, {Skrutskie}, {Slosar}, {Smethurst}, {Sobeck}, {Sodi},
  {Souto}, {Stark}, {Stassun}, {Steinmetz}, {Stello}, {Stermer},
  {Storchi-Bergmann}, {Streblyanska}, {Stringfellow}, {Stutz}, {Su{\'a}rez},
  {Sun}, {Taghizadeh-Popp}, {Talbot}, {Tayar}, {Thakar}, {Theriault}, {Thomas},
  {Thomas}, {Tinker}, {Tojeiro}, {Toledo}, {Tremonti}, {Troup}, {Tuttle},
  {Unda-Sanzana}, {Valentini}, {Vargas-Gonz{\'a}lez}, {Vargas-Maga{\~n}a},
  {V{\'a}zquez-Mata}, {Vivek}, {Wake}, {Wang}, {Weaver}, {Weijmans}, {Wild},
  {Wilson}, {Wilson}, {Wolthuis}, {Wood-Vasey}, {Yan}, {Yang}, {Y{\`e}che},
  {Zamora}, {Zarrouk}, {Zasowski}, {Zhang}, {Zhao}, {Zhao}, {Zheng}, {Zheng},
  {Zhu}, \& {Zou}}]{Ahumada20}
{Ahumada}, R., {Prieto}, C.~A., {Almeida}, A., {et~al.} 2020, \apjs, 249, 3,
  \dodoi{10.3847/1538-4365/ab929e}

\bibitem[{{Betoule} {et~al.}(2014){Betoule}, {Kessler}, {Guy}, {Mosher},
  {Hardin}, {Biswas}, {Astier}, {El-Hage}, {Konig}, {Kuhlmann}, {Marriner},
  {Pain}, {Regnault}, {Balland}, {Bassett}, {Brown}, {Campbell}, {Carlberg},
  {Cellier-Holzem}, {Cinabro}, {Conley}, {D'Andrea}, {DePoy}, {Doi}, {Ellis},
  {Fabbro}, {Filippenko}, {Foley}, {Frieman}, {Fouchez}, {Galbany}, {Goobar},
  {Gupta}, {Hill}, {Hlozek}, {Hogan}, {Hook}, {Howell}, {Jha}, {Le Guillou},
  {Leloudas}, {Lidman}, {Marshall}, {M{\"o}ller}, {Mour{\~a}o}, {Neveu},
  {Nichol}, {Olmstead}, {Palanque-Delabrouille}, {Perlmutter}, {Prieto},
  {Pritchet}, {Richmond}, {Riess}, {Ruhlmann-Kleider}, {Sako}, {Schahmaneche},
  {Schneider}, {Smith}, {Sollerman}, {Sullivan}, {Walton}, \&
  {Wheeler}}]{Betoule2014}
{Betoule}, M., {Kessler}, R., {Guy}, J., {et~al.} 2014, \aap, 568, A22,
  \dodoi{10.1051/0004-6361/201423413}

\bibitem[{{Biswas} {et~al.}(2021){Biswas}, {Goobar}, {Dhawan}, {Schulze},
  {Johansson}, {Bellm}, {Dekany}, {Drake}, {Duev}, {Fremling}, {Graham}, {Kim},
  {Kool}, {Kulkarni}, {Mahabal}, {Perley}, {Rigault}, {Rusholme}, {Sollerman},
  {Shupe}, {Smith}, \& {Walters}}]{Biswas21}
{Biswas}, R., {Goobar}, A., {Dhawan}, S., {et~al.} 2021, \mnras,
  \dodoi{10.1093/mnras/stab2943}

\bibitem[{{Brout} \& {Scolnic}(2021)}]{bs20}
{Brout}, D., \& {Scolnic}, D. 2021, \apj, 909, 26,
  \dodoi{10.3847/1538-4357/abd69b}

\bibitem[{{Brout} {et~al.}(2019{\natexlab{a}}){Brout}, {Scolnic}, {Kessler},
  {D'Andrea}, {Davis}, {Gupta}, {Hinton}, {Kim}, {Lasker}, {Lidman},
  {Macaulay}, {M{\"o}ller}, {Nichol}, {Sako}, {Smith}, {Sullivan}, {Zhang},
  {Andersen}, {Asorey}, {Avelino}, {Bassett}, {Brown}, {Calcino}, {Carollo},
  {Challis}, {Childress}, {Clocchiatti}, {Filippenko}, {Foley}, {Galbany},
  {Glazebrook}, {Hoormann}, {Kasai}, {Kirshner}, {Kuehn}, {Kuhlmann}, {Lewis},
  {Mandel}, {March}, {Miranda}, {Morganson}, {Muthukrishna}, {Nugent},
  {Palmese}, {Pan}, {Sharp}, {Sommer}, {Swann}, {Thomas}, {Tucker}, {Uddin},
  {Wester}, {Abbott}, {Allam}, {Annis}, {Avila}, {Bechtol}, {Bernstein},
  {Bertin}, {Brooks}, {Burke}, {Carnero Rosell}, {Carrasco Kind}, {Carretero},
  {Castander}, {Cunha}, {da Costa}, {Davis}, {De Vicente}, {DePoy}, {Desai},
  {Diehl}, {Doel}, {Drlica-Wagner}, {Eifler}, {Estrada}, {Fernandez},
  {Flaugher}, {Fosalba}, {Frieman}, {Garc{\'\i}a-Bellido}, {Gruen}, {Gruendl},
  {Gutierrez}, {Hartley}, {Hollowood}, {Honscheid}, {Hoyle}, {James}, {Jarvis},
  {Jeltema}, {Krause}, {Lahav}, {Li}, {Lima}, {Maia}, {Marriner}, {Marshall},
  {Martini}, {Menanteau}, {Miller}, {Miquel}, {Ogando}, {Plazas}, {Romer},
  {Roodman}, {Rykoff}, {Sanchez}, {Santiago}, {Scarpine}, {Schubnell},
  {Serrano}, {Sevilla-Noarbe}, {Smith}, {Soares-Santos}, {Sobreira}, {Suchyta},
  {Swanson}, {Tarle}, {Thomas}, {Troxel}, {Tucker}, {Vikram}, {Walker},
  {Zhang}, \& {DES Collaboration}}]{Brout2019}
{Brout}, D., {Scolnic}, D., {Kessler}, R., {et~al.} 2019{\natexlab{a}}, \apj,
  874, 150, \dodoi{10.3847/1538-4357/ab08a0}

\bibitem[{{Brout} {et~al.}(2019{\natexlab{b}}){Brout}, {Sako}, {Scolnic},
  {Kessler}, {D'Andrea}, {Davis}, {Hinton}, {Kim}, {Lasker}, {Macaulay},
  {M{\"o}ller}, {Nichol}, {Smith}, {Sullivan}, {Wolf}, {Allam}, {Bassett},
  {Brown}, {Castander}, {Childress}, {Foley}, {Galbany}, {Herner}, {Kasai},
  {March}, {Morganson}, {Nugent}, {Pan}, {Thomas}, {Tucker}, {Wester},
  {Abbott}, {Annis}, {Avila}, {Bertin}, {Brooks}, {Burke}, {Carnero Rosell},
  {Carrasco Kind}, {Carretero}, {Crocce}, {Cunha}, {da Costa}, {Davis}, {De
  Vicente}, {Desai}, {Diehl}, {Doel}, {Eifler}, {Flaugher}, {Fosalba},
  {Frieman}, {Garc{\'\i}a-Bellido}, {Gaztanaga}, {Gerdes}, {Goldstein},
  {Gruen}, {Gruendl}, {Gschwend}, {Gutierrez}, {Hartley}, {Hollowood},
  {Honscheid}, {James}, {Kuehn}, {Kuropatkin}, {Lahav}, {Li}, {Lima},
  {Marshall}, {Martini}, {Miquel}, {Nord}, {Plazas}, {Roodman}, {Rykoff},
  {Sanchez}, {Scarpine}, {Schindler}, {Schubnell}, {Serrano}, {Sevilla-Noarbe},
  {Soares-Santos}, {Sobreira}, {Suchyta}, {Swanson}, {Tarle}, {Thomas},
  {Tucker}, {Walker}, {Yanny}, {Zhang}, \& {DES COLLABORATION}}]{Brout19}
{Brout}, D., {Sako}, M., {Scolnic}, D., {et~al.} 2019{\natexlab{b}}, \apj, 874,
  106, \dodoi{10.3847/1538-4357/ab06c1}

\bibitem[{{Brown} {et~al.}(2014){Brown}, {Breeveld}, {Holland}, {Kuin}, \&
  {Pritchard}}]{sousa14}
{Brown}, P.~J., {Breeveld}, A.~A., {Holland}, S., {Kuin}, P., \& {Pritchard},
  T. 2014, \apss, 354, 89, \dodoi{10.1007/s10509-014-2059-8}

\bibitem[{{Brownsberger} {et~al.}(2021){Brownsberger}, {Brout}, {Scolnic},
  {Stubbs}, \& {Riess}}]{Brownsberger21}
{Brownsberger}, S., {Brout}, D., {Scolnic}, D., {Stubbs}, C.~W., \& {Riess},
  A.~G. 2021, arXiv e-prints, arXiv:2110.03486.
\newblock \doarXiv{2110.03486}

\bibitem[{{Bruzual} \& {Charlot}(2003)}]{Bruzal03}
{Bruzual}, G., \& {Charlot}, S. 2003, \mnras, 344, 1000,
  \dodoi{10.1046/j.1365-8711.2003.06897.x}

\bibitem[{{Burns} {et~al.}(2018){Burns}, {Parent}, {Phillips}, {Stritzinger},
  {Krisciunas}, {Suntzeff}, {Hsiao}, {Contreras}, {Anais}, {Boldt}, {Busta},
  {Campillay}, {Castell{\'o}n}, {Folatelli}, {Freedman}, {Gonz{\'a}lez},
  {Hamuy}, {Heoflich}, {Krzeminski}, {Madore}, {Morrell}, {Persson}, {Roth},
  {Salgado}, {Ser{\'o}n}, \& {Torres}}]{Burns18}
{Burns}, C.~R., {Parent}, E., {Phillips}, M.~M., {et~al.} 2018, \apj, 869, 56,
  \dodoi{10.3847/1538-4357/aae51c}

\bibitem[{{Burns} {et~al.}(2020){Burns}, {Ashall}, {Contreras}, {Brown},
  {Stritzinger}, {Phillips}, {Flores}, {Suntzeff}, {Hsiao}, {Uddin}, {Simon},
  {Krisciunas}, {Campillay}, {Foley}, {Freedman}, {Galbany}, {Gonz{\'a}lez},
  {Hoeflich}, {Holmbo}, {Kilpatrick}, {Kirshner}, {Morrell},
  {Mu{\~n}oz-Elgueta}, {Piro}, {Rojas-Bravo}, {Sand}, {Vargas-Gonz{\'a}lez},
  {Ulloa}, \& {Vilchez}}]{Burns20}
{Burns}, C.~R., {Ashall}, C., {Contreras}, C., {et~al.} 2020, \apj, 895, 118,
  \dodoi{10.3847/1538-4357/ab8e3e}

\bibitem[{{Carr} {et~al.}(2021){Carr}, {Davis}, {Scolnic}, {Said}, {Brout},
  {Peterson}, \& {Kessler}}]{carr2021pantheon}
{Carr}, A., {Davis}, T.~M., {Scolnic}, D., {et~al.} 2021, arXiv e-prints,
  arXiv:2112.01471.
\newblock \doarXiv{2112.01471}

\bibitem[{{Chabrier}(2003)}]{Chabrier03}
{Chabrier}, G. 2003, \apjl, 586, L133, \dodoi{10.1086/374879}

\bibitem[{{Chambers} \& {et al.}(2017)}]{Chambers17}
{Chambers}, K.~C., \& {et al.} 2017, VizieR Online Data Catalog, II/349

\bibitem[{{Chen} {et~al.}(2020){Chen}, {Dong}, {Kochanek}, {Stanek}, {Post},
  {Stritzinger}, {Prieto}, {Filippenko}, {Kollmeier}, {Katz}, {Tomasella},
  {Bose}, {Benetti}, {Bersier}, {Brink}, {Buckley}, {Cappellaro}, {Elias-Rosa},
  {Fraser}, {Holoien}, {Gromadzki}, {Kankare}, {Lundqvist}, {Mattila},
  {Morrell}, {Shappee}, {Thompson}, \& {Zheng}}]{Chen20}
{Chen}, P., {Dong}, S., {Kochanek}, C.~S., {et~al.} 2020, arXiv e-prints,
  arXiv:2011.02461.
\newblock \doarXiv{2011.02461}

\bibitem[{Conley {et~al.}(2010)Conley, Guy, Sullivan, Regnault, Astier,
  Balland, Basa, Carlberg, Fouchez, Hardin, \& et~al.}]{Conley2010}
Conley, A., Guy, J., Sullivan, M., {et~al.} 2010, The Astrophysical Journal
  Supplement Series, 192, 1, \dodoi{10.1088/0067-0049/192/1/1}

\bibitem[{{Contreras} {et~al.}(2010){Contreras}, {Hamuy}, {Phillips},
  {Folatelli}, {Suntzeff}, {Persson}, {Stritzinger}, {Boldt}, {Gonz{\'a}lez},
  {Krzeminski}, {Morrell}, {Roth}, {Salgado}, {Maureira}, {Burns}, {Freedman},
  {Madore}, {Murphy}, {Wyatt}, {Li}, \& {Filippenko}}]{Contreras2010}
{Contreras}, C., {Hamuy}, M., {Phillips}, M.~M., {et~al.} 2010, \aj, 139, 519,
  \dodoi{10.1088/0004-6256/139/2/519}

\bibitem[{{Currie} {et~al.}(2020){Currie}, {Rubin}, {Aldering}, {Deustua},
  {Fruchter}, \& {Perlmutter}}]{Currie20}
{Currie}, M., {Rubin}, D., {Aldering}, G., {et~al.} 2020, arXiv e-prints,
  arXiv:2007.02458.
\newblock \doarXiv{2007.02458}

\bibitem[{{Dhawan} {et~al.}(2021){Dhawan}, {Goobar}, {Smith}, {Johansson},
  {Rigault}, {Nordin}, {Biswas}, {Goldstein}, {Nugent}, {Kim}, {Miller},
  {Graham}, {Medford}, {Kasliwal}, {Kulkarni}, {Duev}, {Bellm}, {Rosnet},
  {Riddle}, \& {Sollerman}}]{dhawan21}
{Dhawan}, S., {Goobar}, A., {Smith}, M., {et~al.} 2021, \mnras,
  \dodoi{10.1093/mnras/stab3093}

\bibitem[{{Foley} {et~al.}(2018){Foley}, {Scolnic}, {Rest}, {Jha}, {Pan},
  {Riess}, {Challis}, {Chambers}, {Coulter}, {Dettman}, {Foley}, {Fox},
  {Huber}, {Jones}, {Kilpatrick}, {Kirshner}, {Schultz}, {Siebert},
  {Flewelling}, {Gibson}, {Magnier}, {Miller}, {Primak}, {Smartt}, {Smith},
  {Wainscoat}, {Waters}, \& {Willman}}]{Foley2018}
{Foley}, R.~J., {Scolnic}, D., {Rest}, A., {et~al.} 2018, \mnras, 475, 193,
  \dodoi{10.1093/mnras/stx3136}

\bibitem[{{Gall} {et~al.}(2018){Gall}, {Stritzinger}, {Ashall}, {Baron},
  {Burns}, {Hoeflich}, {Hsiao}, {Mazzali}, {Phillips}, {Filippenko},
  {Anderson}, {Benetti}, {Brown}, {Campillay}, {Challis}, {Contreras}, {Elias
  de la Rosa}, {Folatelli}, {Foley}, {Fraser}, {Holmbo}, {Marion}, {Morrell},
  {Pan}, {Pignata}, {Suntzeff}, {Taddia}, {Torres Robledo}, \&
  {Valenti}}]{Gall2018}
{Gall}, C., {Stritzinger}, M.~D., {Ashall}, C., {et~al.} 2018, \aap, 611, A58,
  \dodoi{10.1051/0004-6361/201730886}

\bibitem[{{Ganeshalingam} {et~al.}(2010){Ganeshalingam}, {Li}, {Filippenko},
  {Anderson}, {Foster}, {Gates}, {Griffith}, {Grigsby}, {Joubert}, {Leja},
  {Lowe}, {Macomber}, {Pritchard}, {Thrasher}, \&
  {Winslow}}]{Ganeshalingam2010}
{Ganeshalingam}, M., {Li}, W., {Filippenko}, A.~V., {et~al.} 2010, \apjs, 190,
  418, \dodoi{10.1088/0067-0049/190/2/418}

\bibitem[{{Gilliland} {et~al.}(1999){Gilliland}, {Nugent}, \&
  {Phillips}}]{Gilliland1998}
{Gilliland}, R.~L., {Nugent}, P.~E., \& {Phillips}, M.~M. 1999, \apj, 521, 30,
  \dodoi{10.1086/307549}

\bibitem[{{Gupta} {et~al.}(2016){Gupta}, {Kuhlmann}, {Kovacs}, {Spinka},
  {Kessler}, {Goldstein}, {Liotine}, {Pomian}, {D'Andrea}, {Sullivan},
  {Carretero}, {Castander}, {Nichol}, {Finley}, {Fischer}, {Foley}, {Kim},
  {Papadopoulos}, {Sako}, {Scolnic}, {Smith}, {Tucker}, {Uddin}, {Wolf},
  {Yuan}, {Abbott}, {Abdalla}, {Benoit-L{\'e}vy}, {Bertin}, {Brooks}, {Carnero
  Rosell}, {Carrasco Kind}, {Cunha}, {da Costa}, {Desai}, {Doel}, {Eifler},
  {Evrard}, {Flaugher}, {Fosalba}, {Gazta{\~n}aga}, {Gruen}, {Gruendl},
  {James}, {Kuehn}, {Kuropatkin}, {Maia}, {Marshall}, {Miquel}, {Plazas},
  {Romer}, {S{\'a}nchez}, {Schubnell}, {Sevilla-Noarbe}, {Sobreira}, {Suchyta},
  {Swanson}, {Tarle}, {Walker}, \& {Wester}}]{Gupta16}
{Gupta}, R.~R., {Kuhlmann}, S., {Kovacs}, E., {et~al.} 2016, \aj, 152, 154,
  \dodoi{10.3847/0004-6256/152/6/154}

\bibitem[{{Guy} {et~al.}(2007){Guy}, {Astier}, {Baumont}, {Hardin}, {Pain},
  {Regnault}, {Basa}, {Carlberg}, {Conley}, {Fabbro}, {Fouchez}, {Hook},
  {Howell}, {Perrett}, {Pritchet}, {Rich}, {Sullivan}, {Antilogus}, {Aubourg},
  {Bazin}, {Bronder}, {Filiol}, {Palanque-Delabrouille}, {Ripoche}, \&
  {Ruhlmann-Kleider}}]{Guy07}
{Guy}, J., {Astier}, P., {Baumont}, S., {et~al.} 2007, \aap, 466, 11,
  \dodoi{10.1051/0004-6361:20066930}

\bibitem[{{Hicken} {et~al.}(2009){Hicken}, {Challis}, {Jha}, {Kirshner},
  {Matheson}, {Modjaz}, {Rest}, {Wood-Vasey}, {Bakos}, {Barton}, {Berlind},
  {Bragg}, {Brice{\~n}o}, {Brown}, {Caldwell}, {Calkins}, {Cho}, {Ciupik},
  {Contreras}, {Dendy}, {Dosaj}, {Durham}, {Eriksen}, {Esquerdo}, {Everett},
  {Falco}, {Fernandez}, {Gaba}, {Garnavich}, {Graves}, {Green}, {Groner},
  {Hergenrother}, {Holman}, {Hradecky}, {Huchra}, {Hutchison}, {Jerius},
  {Jordan}, {Kilgard}, {Krauss}, {Luhman}, {Macri}, {Marrone}, {McDowell},
  {McIntosh}, {McNamara}, {Megeath}, {Mochejska}, {Munoz}, {Muzerolle},
  {Naranjo}, {Narayan}, {Pahre}, {Peters}, {Peterson}, {Rines}, {Ripman},
  {Roussanova}, {Schild}, {Sicilia-Aguilar}, {Sokoloski}, {Smalley}, {Smith},
  {Spahr}, {Stanek}, {Barmby}, {Blondin}, {Stubbs}, {Szentgyorgyi}, {Torres},
  {Vaz}, {Vikhlinin}, {Wang}, {Westover}, {Woods}, \& {Zhao}}]{Hicken2009}
{Hicken}, M., {Challis}, P., {Jha}, S., {et~al.} 2009, \apj, 700, 331,
  \dodoi{10.1088/0004-637X/700/1/331}

\bibitem[{{Hicken} {et~al.}(2012){Hicken}, {Challis}, {Kirshner}, {Rest},
  {Cramer}, {Wood-Vasey}, {Bakos}, {Berlind}, {Brown}, {Caldwell}, {Calkins},
  {Currie}, {de Kleer}, {Esquerdo}, {Everett}, {Falco}, {Fernandez},
  {Friedman}, {Groner}, {Hartman}, {Holman}, {Hutchins}, {Keys}, {Kipping},
  {Latham}, {Marion}, {Narayan}, {Pahre}, {Pal}, {Peters}, {Perumpilly},
  {Ripman}, {Sipocz}, {Szentgyorgyi}, {Tang}, {Torres}, {Vaz}, {Wolk}, \&
  {Zezas}}]{Hicken2012}
{Hicken}, M., {Challis}, P., {Kirshner}, R.~P., {et~al.} 2012, \apjs, 200, 12,
  \dodoi{10.1088/0067-0049/200/2/12}

\bibitem[{{Hinton} \& {Brout}(2020)}]{Hinton_and_Brout2020}
{Hinton}, S., \& {Brout}, D. 2020, The Journal of Open Source Software, 5,
  2122, \dodoi{10.21105/joss.02122}

\bibitem[{{Hinton}(2016)}]{Hinton16}
{Hinton}, S.~R. 2016, The Journal of Open Source Software, 1, 00045,
  \dodoi{10.21105/joss.00045}

\bibitem[{{Hounsell} {et~al.}(2018){Hounsell}, {Scolnic}, {Foley}, {Kessler},
  {Miranda}, {Avelino}, {Bohlin}, {Filippenko}, {Frieman}, {Jha}, {Kelly},
  {Kirshner}, {Mandel}, {Rest}, {Riess}, {Rodney}, \& {Strolger}}]{Hounsell18}
{Hounsell}, R., {Scolnic}, D., {Foley}, R.~J., {et~al.} 2018, \apj, 867, 23,
  \dodoi{10.3847/1538-4357/aac08b}

\bibitem[{Hunter(2007)}]{matplotlib}
Hunter, J.~D. 2007, Computing in Science \& Engineering, 9, 90,
  \dodoi{10.1109/MCSE.2007.55}

\bibitem[{{Ilbert} {et~al.}(2006){Ilbert}, {Arnouts}, {McCracken},
  {Bolzonella}, {Bertin}, {Le F{\`e}vre}, {Mellier}, {Zamorani}, {Pell{\`o}},
  {Iovino}, {Tresse}, {Le Brun}, {Bottini}, {Garilli}, {Maccagni}, {Picat},
  {Scaramella}, {Scodeggio}, {Vettolani}, {Zanichelli}, {Adami}, {Bardelli},
  {Cappi}, {Charlot}, {Ciliegi}, {Contini}, {Cucciati}, {Foucaud}, {Franzetti},
  {Gavignaud}, {Guzzo}, {Marano}, {Marinoni}, {Mazure}, {Meneux}, {Merighi},
  {Paltani}, {Pollo}, {Pozzetti}, {Radovich}, {Zucca}, {Bondi}, {Bongiorno},
  {Busarello}, {de La Torre}, {Gregorini}, {Lamareille}, {Mathez}, {Merluzzi},
  {Ripepi}, {Rizzo}, \& {Vergani}}]{lephare06}
{Ilbert}, O., {Arnouts}, S., {McCracken}, H.~J., {et~al.} 2006, \aap, 457, 841,
  \dodoi{10.1051/0004-6361:20065138}

\bibitem[{{Ivezi{\'c}} {et~al.}(2019){Ivezi{\'c}}, {Kahn}, {Tyson}, {Abel},
  {Acosta}, {Allsman}, {Alonso}, {AlSayyad}, {Anderson}, {Andrew}, {Angel},
  {Angeli}, {Ansari}, {Antilogus}, {Araujo}, {Armstrong}, {Arndt}, {Astier},
  {Aubourg}, {Auza}, {Axelrod}, {Bard}, {Barr}, {Barrau}, {Bartlett}, {Bauer},
  {Bauman}, {Baumont}, {Bechtol}, {Bechtol}, {Becker}, {Becla}, {Beldica},
  {Bellavia}, {Bianco}, {Biswas}, {Blanc}, {Blazek}, {Blandford}, {Bloom},
  {Bogart}, {Bond}, {Booth}, {Borgland}, {Borne}, {Bosch}, {Boutigny},
  {Brackett}, {Bradshaw}, {Brandt}, {Brown}, {Bullock}, {Burchat}, {Burke},
  {Cagnoli}, {Calabrese}, {Callahan}, {Callen}, {Carlin}, {Carlson},
  {Chandrasekharan}, {Charles-Emerson}, {Chesley}, {Cheu}, {Chiang}, {Chiang},
  {Chirino}, {Chow}, {Ciardi}, {Claver}, {Cohen-Tanugi}, {Cockrum}, {Coles},
  {Connolly}, {Cook}, {Cooray}, {Covey}, {Cribbs}, {Cui}, {Cutri}, {Daly},
  {Daniel}, {Daruich}, {Daubard}, {Daues}, {Dawson}, {Delgado}, {Dellapenna},
  {de Peyster}, {de Val-Borro}, {Digel}, {Doherty}, {Dubois},
  {Dubois-Felsmann}, {Durech}, {Economou}, {Eifler}, {Eracleous}, {Emmons},
  {Fausti Neto}, {Ferguson}, {Figueroa}, {Fisher-Levine}, {Focke}, {Foss},
  {Frank}, {Freemon}, {Gangler}, {Gawiser}, {Geary}, {Gee}, {Geha}, {Gessner},
  {Gibson}, {Gilmore}, {Glanzman}, {Glick}, {Goldina}, {Goldstein}, {Goodenow},
  {Graham}, {Gressler}, {Gris}, {Guy}, {Guyonnet}, {Haller}, {Harris},
  {Hascall}, {Haupt}, {Hernandez}, {Herrmann}, {Hileman}, {Hoblitt}, {Hodgson},
  {Hogan}, {Howard}, {Huang}, {Huffer}, {Ingraham}, {Innes}, {Jacoby}, {Jain},
  {Jammes}, {Jee}, {Jenness}, {Jernigan}, {Jevremovi{\'c}}, {Johns}, {Johnson},
  {Johnson}, {Jones}, {Juramy-Gilles}, {Juri{\'c}}, {Kalirai}, {Kallivayalil},
  {Kalmbach}, {Kantor}, {Karst}, {Kasliwal}, {Kelly}, {Kessler}, {Kinnison},
  {Kirkby}, {Knox}, {Kotov}, {Krabbendam}, {Krughoff}, {Kub{\'a}nek},
  {Kuczewski}, {Kulkarni}, {Ku}, {Kurita}, {Lage}, {Lambert}, {Lange},
  {Langton}, {Le Guillou}, {Levine}, {Liang}, {Lim}, {Lintott}, {Long},
  {Lopez}, {Lotz}, {Lupton}, {Lust}, {MacArthur}, {Mahabal}, {Mandelbaum},
  {Markiewicz}, {Marsh}, {Marshall}, {Marshall}, {May}, {McKercher}, {McQueen},
  {Meyers}, {Migliore}, {Miller}, {Mills}, {Miraval}, {Moeyens}, {Moolekamp},
  {Monet}, {Moniez}, {Monkewitz}, {Montgomery}, {Morrison}, {Mueller},
  {Muller}, {Mu{\~n}oz Arancibia}, {Neill}, {Newbry}, {Nief}, {Nomerotski},
  {Nordby}, {O'Connor}, {Oliver}, {Olivier}, {Olsen}, {O'Mullane}, {Ortiz},
  {Osier}, {Owen}, {Pain}, {Palecek}, {Parejko}, {Parsons}, {Pease},
  {Peterson}, {Peterson}, {Petravick}, {Libby Petrick}, {Petry},
  {Pierfederici}, {Pietrowicz}, {Pike}, {Pinto}, {Plante}, {Plate}, {Plutchak},
  {Price}, {Prouza}, {Radeka}, {Rajagopal}, {Rasmussen}, {Regnault}, {Reil},
  {Reiss}, {Reuter}, {Ridgway}, {Riot}, {Ritz}, {Robinson}, {Roby}, {Roodman},
  {Rosing}, {Roucelle}, {Rumore}, {Russo}, {Saha}, {Sassolas}, {Schalk},
  {Schellart}, {Schindler}, {Schmidt}, {Schneider}, {Schneider}, {Schoening},
  {Schumacher}, {Schwamb}, {Sebag}, {Selvy}, {Sembroski}, {Seppala}, {Serio},
  {Serrano}, {Shaw}, {Shipsey}, {Sick}, {Silvestri}, {Slater}, {Smith},
  {Smith}, {Sobhani}, {Soldahl}, {Storrie-Lombardi}, {Stover}, {Strauss},
  {Street}, {Stubbs}, {Sullivan}, {Sweeney}, {Swinbank}, {Szalay}, {Takacs},
  {Tether}, {Thaler}, {Thayer}, {Thomas}, {Thornton}, {Thukral}, {Tice},
  {Trilling}, {Turri}, {Van Berg}, {Vanden Berk}, {Vetter}, {Virieux},
  {Vucina}, {Wahl}, {Walkowicz}, {Walsh}, {Walter}, {Wang}, {Wang}, {Warner},
  {Wiecha}, {Willman}, {Winters}, {Wittman}, {Wolff}, {Wood-Vasey}, {Wu},
  {Xin}, {Yoachim}, \& {Zhan}}]{zeljko19}
{Ivezi{\'c}}, {\v{Z}}., {Kahn}, S.~M., {Tyson}, J.~A., {et~al.} 2019, \apj,
  873, 111, \dodoi{10.3847/1538-4357/ab042c}

\bibitem[{{Jha} {et~al.}(2007){Jha}, {Riess}, \& {Kirshner}}]{jha2007}
{Jha}, S., {Riess}, A.~G., \& {Kirshner}, R.~P. 2007, \apj, 659, 122,
  \dodoi{10.1086/512054}

\bibitem[{{Jha} {et~al.}(2006){Jha}, {Kirshner}, {Challis}, {Garnavich},
  {Matheson}, {Soderberg}, {Graves}, {Hicken}, {Alves}, {Arce}, {Balog},
  {Barmby}, {Barton}, {Berlind}, {Bragg}, {Brice{\~n}o}, {Brown}, {Buckley},
  {Caldwell}, {Calkins}, {Carter}, {Concannon}, {Donnelly}, {Eriksen},
  {Fabricant}, {Falco}, {Fiore}, {Garcia}, {G{\'o}mez}, {Grogin}, {Groner},
  {Groot}, {Haisch}, {Hartmann}, {Hergenrother}, {Holman}, {Huchra},
  {Jayawardhana}, {Jerius}, {Kannappan}, {Kim}, {Kleyna}, {Kochanek},
  {Koranyi}, {Krockenberger}, {Lada}, {Luhman}, {Luu}, {Macri}, {Mader},
  {Mahdavi}, {Marengo}, {Marsden}, {McLeod}, {McNamara}, {Megeath}, {Moraru},
  {Mossman}, {Muench}, {Mu{\~n}oz}, {Muzerolle}, {Naranjo}, {Nelson-Patel},
  {Pahre}, {Patten}, {Peters}, {Peters}, {Raymond}, {Rines}, {Schild},
  {Sobczak}, {Spahr}, {Stauffer}, {Stefanik}, {Szentgyorgyi}, {Tollestrup},
  {V{\"a}is{\"a}nen}, {Vikhlinin}, {Wang}, {Willner}, {Wolk}, {Zajac}, {Zhao},
  \& {Stanek}}]{jha06}
{Jha}, S., {Kirshner}, R.~P., {Challis}, P., {et~al.} 2006, \aj, 131, 527,
  \dodoi{10.1086/497989}

\bibitem[{{Jones} {et~al.}(2019){Jones}, {Scolnic}, {Foley}, {Rest}, {Kessler},
  {Challis}, {Chambers}, {Coulter}, {Dettman}, {Foley}, {Huber}, {Jha},
  {Johnson}, {Kilpatrick}, {Kirshner}, {Manuel}, {Narayan}, {Pan}, {Riess},
  {Schultz}, {Siebert}, {Berger}, {Chornock}, {Flewelling}, {Magnier},
  {Smartt}, {Smith}, {Wainscoat}, {Waters}, \& {Willman}}]{Jones19}
{Jones}, D.~O., {Scolnic}, D.~M., {Foley}, R.~J., {et~al.} 2019, \apj, 881, 19,
  \dodoi{10.3847/1538-4357/ab2bec}

\bibitem[{{Jones} {et~al.}(2021){Jones}, {Foley}, {Narayan}, {Hjorth}, {Huber},
  {Aleo}, {Alexander}, {Angus}, {Auchettl}, {Baldassare}, {Bruun}, {Chambers},
  {Chatterjee}, {Coppejans}, {Coulter}, {DeMarchi}, {Dimitriadis}, {Drout},
  {Engel}, {French}, {Gagliano}, {Gall}, {Hung}, {Izzo}, {Jacobson-Gal{\'a}n},
  {Kilpatrick}, {Korhonen}, {Margutti}, {Raimundo}, {Ramirez-Ruiz}, {Rest},
  {Rojas-Bravo}, {Siebert}, {Smartt}, {Smith}, {Terreran}, {Wang}, {Wojtak},
  {Agnello}, {Ansari}, {Arendse}, {Baldeschi}, {Blanchard}, {Brethauer},
  {Bright}, {Brown}, {de Boer}, {Dodd}, {Fairlamb}, {Grillo}, {Hajela}, {Hede},
  {Kolborg}, {Law-Smith}, {Lin}, {Magnier}, {Malanchev}, {Matthews}, {Mockler},
  {Muthukrishna}, {Pan}, {Pfister}, {Ramanah}, {Rest}, {Sarangi},
  {Schr{\o}der}, {Stauffer}, {Stroh}, {Taggart}, {Tinyanont}, {Wainscoat}, \&
  {Young Supernova Experiment}}]{jones21}
{Jones}, D.~O., {Foley}, R.~J., {Narayan}, G., {et~al.} 2021, \apj, 908, 143,
  \dodoi{10.3847/1538-4357/abd7f5}

\bibitem[{{Kawabata} {et~al.}(2020){Kawabata}, {Maeda}, {Yamanaka}, {Nakaoka},
  {Kawabata}, {Adachi}, {Akitaya}, {Burgaz}, {Hanayama}, {Horiuchi},
  {Hosokawa}, {Iida}, {Imazato}, {Isogai}, {Jiang}, {Katoh}, {Kimura}, {Kino},
  {Kuroda}, {Maehara}, {Matsubayashi}, {Morihana}, {Murata}, {Nagao}, {Niwano},
  {Nogami}, {Oeda}, {Ono}, {Onozato}, {Otsuka}, {Saito}, {Sasada}, {Shiraishi},
  {Sugiyama}, {Taguchi}, {Takahashi}, {Takagi}, {Takagi}, {Takayama}, {Tozuka},
  \& {Sekiguchi}}]{Kawabata20}
{Kawabata}, M., {Maeda}, K., {Yamanaka}, M., {et~al.} 2020, \apj, 893, 143,
  \dodoi{10.3847/1538-4357/ab8236}

\bibitem[{{Kelly} {et~al.}(2010){Kelly}, {Hicken}, {Burke}, {Mandel}, \&
  {Kirshner}}]{Kelly10}
{Kelly}, P.~L., {Hicken}, M., {Burke}, D.~L., {Mandel}, K.~S., \& {Kirshner},
  R.~P. 2010, \apj, 715, 743, \dodoi{10.1088/0004-637X/715/2/743}

\bibitem[{{Kessler} {et~al.}(2009){Kessler}, {Bernstein}, {Cinabro}, {Dilday},
  {Frieman}, {Jha}, {Kuhlmann}, {Miknaitis}, {Sako}, {Taylor}, \&
  {Vanderplas}}]{Kessler2009}
{Kessler}, R., {Bernstein}, J.~P., {Cinabro}, D., {et~al.} 2009, \pasp, 121,
  1028, \dodoi{10.1086/605984}

\bibitem[{{Krisciunas} {et~al.}(2017{\natexlab{a}}){Krisciunas}, {Suntzeff},
  {Espinoza}, {Gonzalez}, {Miranda}, \& {Sanhueza}}]{Kris2017_df}
{Krisciunas}, K., {Suntzeff}, N.~B., {Espinoza}, J., {et~al.}
  2017{\natexlab{a}}, Research Notes of the American Astronomical Society, 1,
  36, \dodoi{10.3847/2515-5172/aa9f18}

\bibitem[{{Krisciunas} {et~al.}(2017{\natexlab{b}}){Krisciunas}, {Contreras},
  {Burns}, {Phillips}, {Stritzinger}, {Morrell}, {Hamuy}, {Anais}, {Boldt},
  {Busta}, {Campillay}, {Castell{\'o}n}, {Folatelli}, {Freedman},
  {Gonz{\'a}lez}, {Hsiao}, {Krzeminski}, {Persson}, {Roth}, {Salgado},
  {Ser{\'o}n}, {Suntzeff}, {Torres}, {Filippenko}, {Li}, {Madore}, {DePoy},
  {Marshall}, {Rheault}, \& {Villanueva}}]{Krisciunas2017}
{Krisciunas}, K., {Contreras}, C., {Burns}, C.~R., {et~al.} 2017{\natexlab{b}},
  \aj, 154, 211, \dodoi{10.3847/1538-3881/aa8df0}

\bibitem[{{Lampeitl} {et~al.}(2010){Lampeitl}, {Smith}, {Nichol}, {Bassett},
  {Cinabro}, {Dilday}, {Foley}, {Frieman}, {Garnavich}, {Goobar}, {Im}, {Jha},
  {Marriner}, {Miquel}, {Nordin}, {{\"O}stman}, {Riess}, {Sako}, {Schneider},
  {Sollerman}, \& {Stritzinger}}]{lampeitl10}
{Lampeitl}, H., {Smith}, M., {Nichol}, R.~C., {et~al.} 2010, \apj, 722, 566,
  \dodoi{10.1088/0004-637X/722/1/566}

\bibitem[{{Martin} {et~al.}(2005){Martin}, {Fanson}, {Schiminovich},
  {Morrissey}, {Friedman}, {Barlow}, {Conrow}, {Grange}, {Jelinsky},
  {Milliard}, {Siegmund}, {Bianchi}, {Byun}, {Donas}, {Forster}, {Heckman},
  {Lee}, {Madore}, {Malina}, {Neff}, {Rich}, {Small}, {Surber}, {Szalay},
  {Welsh}, \& {Wyder}}]{Galex05}
{Martin}, D.~C., {Fanson}, J., {Schiminovich}, D., {et~al.} 2005, \apjl, 619,
  L1, \dodoi{10.1086/426387}

\bibitem[{{Milne} {et~al.}(2010){Milne}, {Brown}, {Roming}, {Holland},
  {Immler}, {Filippenko}, {Ganeshalingam}, {Li}, {Stritzinger}, {Phillips},
  {Hicken}, {Kirshner}, {Challis}, {Mazzali}, {Schmidt}, {Bufano}, {Gehrels},
  \& {Vanden Berk}}]{Milne10}
{Milne}, P.~A., {Brown}, P.~J., {Roming}, P. W.~A., {et~al.} 2010, \apj, 721,
  1627, \dodoi{10.1088/0004-637X/721/2/1627}

\bibitem[{Oliphant(2006)}]{numpy}
Oliphant, T.~E. 2006, A guide to NumPy, Vol.~1 (Trelgol Publishing USA)

\bibitem[{{Onken} {et~al.}(2019){Onken}, {Wolf}, {Bessell}, {Chang}, {Da
  Costa}, {Luvaul}, {Mackey}, {Schmidt}, \& {Shao}}]{Onken19}
{Onken}, C.~A., {Wolf}, C., {Bessell}, M.~S., {et~al.} 2019, \pasa, 36, e033,
  \dodoi{10.1017/pasa.2019.27}

\bibitem[{{Perlmutter} {et~al.}(1999){Perlmutter}, {Aldering}, {Goldhaber},
  {Knop}, {Nugent}, {Castro}, {Deustua}, {Fabbro}, {Goobar}, {Groom}, {Hook},
  {Kim}, {Kim}, {Lee}, {Nunes}, {Pain}, {Pennypacker}, {Quimby}, {Lidman},
  {Ellis}, {Irwin}, {McMahon}, {Ruiz-Lapuente}, {Walton}, {Schaefer}, {Boyle},
  {Filippenko}, {Matheson}, {Fruchter}, {Panagia}, {Newberg}, {Couch}, \&
  {Project}}]{Perlmutter1999}
{Perlmutter}, S., {Aldering}, G., {Goldhaber}, G., {et~al.} 1999, \apj, 517,
  565, \dodoi{10.1086/307221}

\bibitem[{{Peterson} {et~al.}(2021){Peterson}, {Kenworthy}, {Scolnic}, {Riess},
  {Brout}, {Carr}, {Courtois}, {Davis}, {Dwomoh}, {Jones}, {Popovic}, {Rose},
  \& {Said}}]{Peterson2021}
{Peterson}, E.~R., {Kenworthy}, W.~D., {Scolnic}, D., {et~al.} 2021, arXiv
  e-prints, arXiv:2110.03487.
\newblock \doarXiv{2110.03487}

\bibitem[{{Popovic} {et~al.}(2021){Popovic}, {Brout}, {Kessler}, {Scolnic}, \&
  {Lu}}]{Popovic21}
{Popovic}, B., {Brout}, D., {Kessler}, R., {Scolnic}, D., \& {Lu}, L. 2021,
  \apj, 913, 49, \dodoi{10.3847/1538-4357/abf14f}

\bibitem[{Price-Whelan {et~al.}(2018)Price-Whelan, Sip{\H{o}}cz, G{\"u}nther,
  Lim, Crawford, Conseil, Shupe, Craig, Dencheva, Ginsburg, {et~al.}}]{astropy}
Price-Whelan, A.~M., Sip{\H{o}}cz, B., G{\"u}nther, H., {et~al.} 2018, The
  Astronomical Journal, 156, 123

\bibitem[{{Riess} {et~al.}(1998){Riess}, {Filippenko}, {Challis},
  {Clocchiatti}, {Diercks}, {Garnavich}, {Gilliland}, {Hogan}, {Jha},
  {Kirshner}, {Leibundgut}, {Phillips}, {Reiss}, {Schmidt}, {Schommer},
  {Smith}, {Spyromilio}, {Stubbs}, {Suntzeff}, \& {Tonry}}]{Riess1998}
{Riess}, A.~G., {Filippenko}, A.~V., {Challis}, P., {et~al.} 1998, \aj, 116,
  1009, \dodoi{10.1086/300499}

\bibitem[{{Riess} {et~al.}(1999){Riess}, {Kirshner}, {Schmidt}, {Jha},
  {Challis}, {Garnavich}, {Esin}, {Carpenter}, {Grashius}, {Schild}, {Berlind},
  {Huchra}, {Prosser}, {Falco}, {Benson}, {Brice{\~n}o}, {Brown}, {Caldwell},
  {dell'Antonio}, {Filippenko}, {Goodman}, {Grogin}, {Groner}, {Hughes},
  {Green}, {Jansen}, {Kleyna}, {Luu}, {Macri}, {McLeod}, {McLeod}, {McNamara},
  {McLean}, {Milone}, {Mohr}, {Moraru}, {Peng}, {Peters}, {Prestwich},
  {Stanek}, {Szentgyorgyi}, \& {Zhao}}]{Riess99}
{Riess}, A.~G., {Kirshner}, R.~P., {Schmidt}, B.~P., {et~al.} 1999, \aj, 117,
  707, \dodoi{10.1086/300738}

\bibitem[{{Riess} {et~al.}(2001){Riess}, {Nugent}, {Gilliland}, {Schmidt},
  {Tonry}, {Dickinson}, {Thompson}, {Budav{\'a}ri}, {Casertano}, {Evans},
  {Filippenko}, {Livio}, {Sanders}, {Shapley}, {Spinrad}, {Steidel}, {Stern},
  {Surace}, \& {Veilleux}}]{Riess2001}
{Riess}, A.~G., {Nugent}, P.~E., {Gilliland}, R.~L., {et~al.} 2001, \apj, 560,
  49, \dodoi{10.1086/322348}

\bibitem[{{Riess} {et~al.}(2004){Riess}, {Strolger}, {Tonry}, {Casertano},
  {Ferguson}, {Mobasher}, {Challis}, {Filippenko}, {Jha}, {Li}, {Chornock},
  {Kirshner}, {Leibundgut}, {Dickinson}, {Livio}, {Giavalisco}, {Steidel},
  {Ben{\'\i}tez}, \& {Tsvetanov}}]{Riess2004}
{Riess}, A.~G., {Strolger}, L.-G., {Tonry}, J., {et~al.} 2004, \apj, 607, 665,
  \dodoi{10.1086/383612}

\bibitem[{{Riess} {et~al.}(2007){Riess}, {Strolger}, {Casertano}, {Ferguson},
  {Mobasher}, {Gold}, {Challis}, {Filippenko}, {Jha}, {Li}, {Tonry}, {Foley},
  {Kirshner}, {Dickinson}, {MacDonald}, {Eisenstein}, {Livio}, {Younger}, {Xu},
  {Dahl{\'e}n}, \& {Stern}}]{Riess2007}
{Riess}, A.~G., {Strolger}, L.-G., {Casertano}, S., {et~al.} 2007, \apj, 659,
  98, \dodoi{10.1086/510378}

\bibitem[{{Riess} {et~al.}(2016){Riess}, {Macri}, {Hoffmann}, {Scolnic},
  {Casertano}, {Filippenko}, {Tucker}, {Reid}, {Jones}, {Silverman},
  {Chornock}, {Challis}, {Yuan}, {Brown}, \& {Foley}}]{Riess2016}
{Riess}, A.~G., {Macri}, L.~M., {Hoffmann}, S.~L., {et~al.} 2016, \apj, 826,
  56, \dodoi{10.3847/0004-637X/826/1/56}

\bibitem[{{Riess} {et~al.}(2018){Riess}, {Rodney}, {Scolnic}, {Shafer},
  {Strolger}, {Ferguson}, {Postman}, {Graur}, {Maoz}, {Jha}, {Mobasher},
  {Casertano}, {Hayden}, {Molino}, {Hjorth}, {Garnavich}, {Jones}, {Kirshner},
  {Koekemoer}, {Grogin}, {Brammer}, {Hemmati}, {Dickinson}, {Challis}, {Wolff},
  {Clubb}, {Filippenko}, {Nayyeri}, {U}, {Koo}, {Faber}, {Kocevski}, {Bradley},
  \& {Coe}}]{Riess2018}
{Riess}, A.~G., {Rodney}, S.~A., {Scolnic}, D.~M., {et~al.} 2018, \apj, 853,
  126, \dodoi{10.3847/1538-4357/aaa5a9}

\bibitem[{{Sako} {et~al.}(2018){Sako}, {Bassett}, {Becker}, {Brown},
  {Campbell}, {Wolf}, {Cinabro}, {D'Andrea}, {Dawson}, {DeJongh}, {Depoy},
  {Dilday}, {Doi}, {Filippenko}, {Fischer}, {Foley}, {Frieman}, {Galbany},
  {Garnavich}, {Goobar}, {Gupta}, {Hill}, {Hayden}, {Hlozek}, {Holtzman},
  {Hopp}, {Jha}, {Kessler}, {Kollatschny}, {Leloudas}, {Marriner}, {Marshall},
  {Miquel}, {Morokuma}, {Mosher}, {Nichol}, {Nordin}, {Olmstead}, {{\"O}stman},
  {Prieto}, {Richmond}, {Romani}, {Sollerman}, {Stritzinger}, {Schneider},
  {Smith}, {Wheeler}, {Yasuda}, \& {Zheng}}]{Sako2018}
{Sako}, M., {Bassett}, B., {Becker}, A.~C., {et~al.} 2018, \pasp, 130, 064002,
  \dodoi{10.1088/1538-3873/aab4e0}

\bibitem[{{Schlafly} \& {Finkbeiner}(2011)}]{Schlafly11}
{Schlafly}, E.~F., \& {Finkbeiner}, D.~P. 2011, \apj, 737, 103,
  \dodoi{10.1088/0004-637X/737/2/103}

\bibitem[{{Scolnic} {et~al.}(2015){Scolnic}, {Casertano}, {Riess}, {Rest},
  {Schlafly}, {Foley}, {Finkbeiner}, {Tang}, {Burgett}, {Chambers}, {Draper},
  {Flewelling}, {Hodapp}, {Huber}, {Kaiser}, {Kudritzki}, {Magnier},
  {Metcalfe}, \& {Stubbs}}]{Scolnic2015}
{Scolnic}, D., {Casertano}, S., {Riess}, A., {et~al.} 2015, \apj, 815, 117,
  \dodoi{10.1088/0004-637X/815/2/117}

\bibitem[{{Scolnic} {et~al.}(2019){Scolnic}, {Perlmutter}, {Aldering}, {Brout},
  {Davis}, {Filippenko}, {Foley}, {Hlo{\v{z}}ek}, {Hounsell}, {Jones}, {Kelly},
  {Rubin}, {Riess}, {Rodney}, {Roberts-Pierel}, {Wang}, {Asorey}, {Avelino},
  {Bavdhankar}, {Brown}, {Challinor}, {Balland}, {Cooray}, {Dhawan},
  {Dimitriadis}, {Dvorkin}, {Guy}, {Handley}, {Keeley}, {Kneib}, {L'Huillier},
  {Lattanzi}, {Mandel}, {Mertens}, {Rigault}, {Motloch}, {Mukherjee},
  {Narayan}, {Nomerotski}, {Page}, {Pogosian}, {Puglisi}, {Raveri}, {Regnault},
  {Rest}, {Rojas-Bravo}, {Sako}, {Shi}, {Sridhar}, {Suzuki}, {Tsai},
  {Wood-Vasey}, {Copin}, {Zhao}, \& {Zhu}}]{Scolnic_Decadal}
{Scolnic}, D., {Perlmutter}, S., {Aldering}, G., {et~al.} 2019, Astro2020:
  Decadal Survey on Astronomy and Astrophysics, 2020, 270.
\newblock \doarXiv{1903.05128}

\bibitem[{{Scolnic} {et~al.}(2020){Scolnic}, {Smith}, {Massiah}, {Wiseman},
  {Brout}, {Kessler}, {Davis}, {Foley}, {Galbany}, {Hinton}, {Hounsell},
  {Kelsey}, {Lidman}, {Macaulay}, {Morgan}, {Nichol}, {M{\"o}ller}, {Popovic},
  {Sako}, {Sullivan}, {Thomas}, {Tucker}, {Abbott}, {Aguena}, {Allam}, {Annis},
  {Avila}, {Bechtol}, {Bertin}, {Brooks}, {Burke}, {Rosell}, {Carollo}, {Kind},
  {Carretero}, {Costanzi}, {da Costa}, {De Vicente}, {Desai}, {Diehl}, {Doel},
  {Drlica-Wagner}, {Eckert}, {Eifler}, {Everett}, {Flaugher}, {Fosalba},
  {Frieman}, {Garc{\'\i}a-Bellido}, {Gaztanaga}, {Gerdes}, {Glazebrook},
  {Gruen}, {Gruendl}, {Gschwend}, {Gutierrez}, {Hartley}, {Hollowood},
  {Honscheid}, {James}, {Kuehn}, {Kuropatkin}, {Lewis}, {Li}, {Lima}, {Maia},
  {Marshall}, {Menanteau}, {Miquel}, {Palmese}, {Paz-Chinch{\'o}n}, {Plazas},
  {Pursiainen}, {Sanchez}, {Scarpine}, {Schubnell}, {Serrano},
  {Sevilla-Noarbe}, {Sommer}, {Suchyta}, {Swanson}, {Tarle}, {Varga}, {Walker},
  {Wilkinson}, \& {DES Collaboration}}]{Scolnic20}
{Scolnic}, D., {Smith}, M., {Massiah}, A., {et~al.} 2020, \apjl, 896, L13,
  \dodoi{10.3847/2041-8213/ab8735}

\bibitem[{{Scolnic} {et~al.}(2018){Scolnic}, {Jones}, {Rest}, {Pan},
  {Chornock}, {Foley}, {Huber}, {Kessler}, {Narayan}, {Riess}, {Rodney},
  {Berger}, {Brout}, {Challis}, {Drout}, {Finkbeiner}, {Lunnan}, {Kirshner},
  {Sand ers}, {Schlafly}, {Smartt}, {Stubbs}, {Tonry}, {Wood-Vasey}, {Foley},
  {Hand}, {Johnson}, {Burgett}, {Chambers}, {Draper}, {Hodapp}, {Kaiser},
  {Kudritzki}, {Magnier}, {Metcalfe}, {Bresolin}, {Gall}, {Kotak}, {McCrum}, \&
  {Smith}}]{Scolnic2018}
{Scolnic}, D.~M., {Jones}, D.~O., {Rest}, A., {et~al.} 2018, \apj, 859, 101,
  \dodoi{10.3847/1538-4357/aab9bb}

\bibitem[{{Skrutskie} {et~al.}(2006){Skrutskie}, {Cutri}, {Stiening},
  {Weinberg}, {Schneider}, {Carpenter}, {Beichman}, {Capps}, {Chester},
  {Elias}, {Huchra}, {Liebert}, {Lonsdale}, {Monet}, {Price}, {Seitzer},
  {Jarrett}, {Kirkpatrick}, {Gizis}, {Howard}, {Evans}, {Fowler}, {Fullmer},
  {Hurt}, {Light}, {Kopan}, {Marsh}, {McCallon}, {Tam}, {Van Dyk}, \&
  {Wheelock}}]{2mass06}
{Skrutskie}, M.~F., {Cutri}, R.~M., {Stiening}, R., {et~al.} 2006, \aj, 131,
  1163, \dodoi{10.1086/498708}

\bibitem[{{Smith} {et~al.}(2020){Smith}, {D'Andrea}, {Sullivan}, {M{\"o}ller},
  {Nichol}, {Thomas}, {Kim}, {Sako}, {Castander}, {Filippenko}, {Foley},
  {Galbany}, {Gonz{\'a}lez-Gait{\'a}n}, {Kasai}, {Kirshner}, {Lidman},
  {Scolnic}, {Brout}, {Davis}, {Gupta}, {Hinton}, {Kessler}, {Lasker},
  {Macaulay}, {Wolf}, {Zhang}, {Asorey}, {Avelino}, {Bassett}, {Calcino},
  {Carollo}, {Casas}, {Challis}, {Childress}, {Clocchiatti}, {Crawford},
  {Frohmaier}, {Glazebrook}, {Goldstein}, {Graham}, {Hoormann}, {Kuehn},
  {Lewis}, {Mandel}, {Morganson}, {Muthukrishna}, {Nugent}, {Pan},
  {Pursiainen}, {Sharp}, {Sommer}, {Swann}, {Thomas}, {Tucker}, {Uddin},
  {Wiseman}, {Zheng}, {Abbott}, {Annis}, {Avila}, {Bechtol}, {Bernstein},
  {Bertin}, {Brooks}, {Burke}, {Carnero Rosell}, {Carrasco Kind}, {Carretero},
  {Cunha}, {da Costa}, {Davis}, {De Vicente}, {Diehl}, {Eifler}, {Estrada},
  {Frieman}, {Garc{\'\i}a-Bellido}, {Gaztanaga}, {Gerdes}, {Gruen}, {Gruendl},
  {Gschwend}, {Gutierrez}, {Hartley}, {Hollowood}, {Honscheid}, {Hoyle},
  {James}, {Johnson}, {Johnson}, {Kuropatkin}, {Li}, {Lima}, {Maia}, {March},
  {Marshall}, {Martini}, {Menanteau}, {Miller}, {Miquel}, {Neilsen}, {Ogando},
  {Plazas}, {Romer}, {Sanchez}, {Scarpine}, {Schubnell}, {Serrano},
  {Sevilla-Noarbe}, {Soares-Santos}, {Sobreira}, {Suchyta}, {Tarle}, {Tucker},
  \& {Wester}}]{Smith2020}
{Smith}, M., {D'Andrea}, C.~B., {Sullivan}, M., {et~al.} 2020, \aj, 160, 267,
  \dodoi{10.3847/1538-3881/abc01b}

\bibitem[{{Stahl} {et~al.}(2019){Stahl}, {Zheng}, {de Jaeger}, {Filippenko},
  {Bigley}, {Blanchard}, {Blanchard}, {Brink}, {Cargill}, {Casper}, {Channa},
  {Choi}, {Choksi}, {Chu}, {Clubb}, {Cohen}, {Ellison}, {Falcon}, {Fazeli},
  {Fuller}, {Ganeshalingam}, {Gates}, {Gould}, {Halevi}, {Hayakawa},
  {Hestenes}, {Jeffers}, {Joubert}, {Kandrashoff}, {Kim}, {Kim}, {Kislak},
  {Kleiser}, {Kong}, {de Kouchkovsky}, {Krishnan}, {Kumar}, {Leja}, {Leonard},
  {Li}, {Li}, {Lu}, {Mason}, {Molloy}, {Pina}, {Rex}, {Ross}, {Stegman},
  {Tang}, {Thrasher}, {Wang}, {Wilkins}, {Yuk}, {Yunus}, \&
  {Zhang}}]{Stahl2019}
{Stahl}, B.~E., {Zheng}, W., {de Jaeger}, T., {et~al.} 2019, \mnras, 490, 3882,
  \dodoi{10.1093/mnras/stz2742}

\bibitem[{{Stritzinger} {et~al.}(2010){Stritzinger}, {Burns}, {Phillips},
  {Folatelli}, {Krisciunas}, {Kattner}, {Persson}, {Boldt}, {Campillay},
  {Contreras}, {Krzeminski}, {Morrell}, {Salgado}, {Freedman}, {Hamuy},
  {Madore}, {Roth}, \& {Suntzeff}}]{Stritzinger10}
{Stritzinger}, M., {Burns}, C.~R., {Phillips}, M.~M., {et~al.} 2010, \aj, 140,
  2036, \dodoi{10.1088/0004-6256/140/6/2036}

\bibitem[{{Stritzinger} {et~al.}(2011){Stritzinger}, {Phillips}, {Boldt},
  {Burns}, {Campillay}, {Contreras}, {Gonzalez}, {Folatelli}, {Morrell},
  {Krzeminski}, {Roth}, {Salgado}, {DePoy}, {Hamuy}, {Freedman}, {Madore},
  {Marshall}, {Persson}, {Rheault}, {Suntzeff}, {Villanueva}, {Li}, \&
  {Filippenko}}]{Stritzinger2011}
{Stritzinger}, M.~D., {Phillips}, M.~M., {Boldt}, L.~N., {et~al.} 2011, \aj,
  142, 156, \dodoi{10.1088/0004-6256/142/5/156}

\bibitem[{{Sullivan} {et~al.}(2006){Sullivan}, {Le Borgne}, {Pritchet},
  {Hodsman}, {Neill}, {Howell}, {Carlberg}, {Astier}, {Aubourg}, {Balam},
  {Basa}, {Conley}, {Fabbro}, {Fouchez}, {Guy}, {Hook}, {Pain},
  {Palanque-Delabrouille}, {Perrett}, {Regnault}, {Rich}, {Taillet}, {Baumont},
  {Bronder}, {Ellis}, {Filiol}, {Lusset}, {Perlmutter}, {Ripoche}, \&
  {Tao}}]{Sullivan06}
{Sullivan}, M., {Le Borgne}, D., {Pritchet}, C.~J., {et~al.} 2006, \apj, 648,
  868, \dodoi{10.1086/506137}

\bibitem[{{Sullivan} {et~al.}(2010){Sullivan}, {Conley}, {Howell}, {Neill},
  {Astier}, {Balland}, {Basa}, {Carlberg}, {Fouchez}, {Guy}, {Hardin}, {Hook},
  {Pain}, {Palanque-Delabrouille}, {Perrett}, {Pritchet}, {Regnault}, {Rich},
  {Ruhlmann-Kleider}, {Baumont}, {Hsiao}, {Kronborg}, {Lidman}, {Perlmutter},
  \& {Walker}}]{Sullivan10}
{Sullivan}, M., {Conley}, A., {Howell}, D.~A., {et~al.} 2010, \mnras, 406, 782,
  \dodoi{10.1111/j.1365-2966.2010.16731.x}

\bibitem[{{Suzuki} {et~al.}(2012){Suzuki}, {Rubin}, {Lidman}, {Aldering},
  {Amanullah}, {Barbary}, {Barrientos}, {Botyanszki}, {Brodwin}, {Connolly},
  {Dawson}, {Dey}, {Doi}, {Donahue}, {Deustua}, {Eisenhardt}, {Ellingson},
  {Faccioli}, {Fadeyev}, {Fakhouri}, {Fruchter}, {Gilbank}, {Gladders},
  {Goldhaber}, {Gonzalez}, {Goobar}, {Gude}, {Hattori}, {Hoekstra}, {Hsiao},
  {Huang}, {Ihara}, {Jee}, {Johnston}, {Kashikawa}, {Koester}, {Konishi},
  {Kowalski}, {Linder}, {Lubin}, {Melbourne}, {Meyers}, {Morokuma}, {Munshi},
  {Mullis}, {Oda}, {Panagia}, {Perlmutter}, {Postman}, {Pritchard}, {Rhodes},
  {Ripoche}, {Rosati}, {Schlegel}, {Spadafora}, {Stanford}, {Stanishev},
  {Stern}, {Strovink}, {Takanashi}, {Tokita}, {Wagner}, {Wang}, {Yasuda},
  {Yee}, \& {Supernova Cosmology Project}}]{Suzuki2012}
{Suzuki}, N., {Rubin}, D., {Lidman}, C., {et~al.} 2012, \apj, 746, 85,
  \dodoi{10.1088/0004-637X/746/1/85}

\bibitem[{{The LSST Dark Energy Science Collaboration} {et~al.}(2018){The LSST
  Dark Energy Science Collaboration}, {Mandelbaum}, {Eifler}, {Hlo{\v{z}}ek},
  {Collett}, {Gawiser}, {Scolnic}, {Alonso}, {Awan}, {Biswas}, {Blazek},
  {Burchat}, {Chisari}, {Dell'Antonio}, {Digel}, {Frieman}, {Goldstein},
  {Hook}, {Ivezi{\'c}}, {Kahn}, {Kamath}, {Kirkby}, {Kitching}, {Krause},
  {Leget}, {Marshall}, {Meyers}, {Miyatake}, {Newman}, {Nichol}, {Rykoff},
  {Sanchez}, {Slosar}, {Sullivan}, \& {Troxel}}]{Mandelbaum18}
{The LSST Dark Energy Science Collaboration}, {Mandelbaum}, R., {Eifler}, T.,
  {et~al.} 2018, arXiv e-prints, arXiv:1809.01669.
\newblock \doarXiv{1809.01669}

\bibitem[{{Tripp}(1998)}]{Tripp}
{Tripp}, R. 1998, \aap, 331, 815

\bibitem[{{Tsvetkov} \& {Elenin}(2010)}]{Tsvetkov10}
{Tsvetkov}, D.~Y., \& {Elenin}, L. 2010, Peremennye Zvezdy, 30, 2.
\newblock \doarXiv{1003.2558}

\bibitem[{Virtanen {et~al.}(2020)Virtanen, Gommers, Oliphant, Haberland, Reddy,
  Cournapeau, Burovski, Peterson, {Weckesser}, {Bright}, {van der Walt},
  {Brett}, {Wilson}, {Jarrod Millman}, {Mayorov}, {Nelson}, {Jones}, {Kern},
  {Larson}, {Carey}, {Polat}, {Feng}, {Moore}, {Vand erPlas}, {Laxalde},
  {Perktold}, {Cimrman}, {Henriksen}, {Quintero}, {Harris}, {Archibald},
  {Ribeiro}, {Pedregosa}, {van Mulbregt}, \& {Contributors}}]{scipy}
Virtanen, P., Gommers, R., Oliphant, T.~E., {et~al.} 2020, Nature Methods

\bibitem[{{Wee} {et~al.}(2018){Wee}, {Chakraborty}, {Wang}, \&
  {Penprase}}]{Wee2018}
{Wee}, J., {Chakraborty}, N., {Wang}, J., \& {Penprase}, B.~E. 2018, \apj, 863,
  90, \dodoi{10.3847/1538-4357/aacd4e}

\bibitem[{{Zhang} {et~al.}(2010){Zhang}, {Wang}, {Li}, {Filippenko}, {Wang},
  {Zhou}, {Brown}, {Silverman}, {Steele}, {Ganeshalingam}, {Li}, {Deng}, {Li},
  {Qiu}, {Zhai}, \& {Shang}}]{Zhang10}
{Zhang}, T., {Wang}, X., {Li}, W., {et~al.} 2010, \pasp, 122, 1,
  \dodoi{10.1086/649851}

\end{thebibliography}
\bibliographystyle{aasjournal}
\section{Acknowledgements}
 D.S. is supported by Department of Energy grant DE-SC0010007 and the David and Lucile Packard Foundation. This manuscript is based upon work supported by the National Aeronautics and Space Administration (NASA) under Contracts NNG16PJ34C and NNG17PX03C issued through the {\it WFIRST} Science Investigation Teams Program.  
D.S.\ and R.J.F.\ are supported in part by NASA grant 14-WPS14-0048.
T.M.D.\ and A.C.\ are supported by an Australian Research Council Laureate Fellowship, FL180100168.  D.O.J.\ is supported by NASA through Hubble Fellowship grant HF2-51462.001 awarded by the Space Telescope Science Institute (STScI), which is operated by the Association of Universities for Research in Astronomy, Inc., for NASA, under contract NAS5-26555. P.J.B.'s work is supported by NASA ADAP grant 80NSSC20K0456:  ``SOUSA's Sequel: Improving Standard Candles by Improving UV Calibration.''  The Pan-STARRS1 Surveys (PS1) and the PS1 public science archive have been made possible through contributions by the Institute for Astronomy, the University of Hawaii, the Pan-STARRS Project Office, the Max-Planck Society and its participating institutes, the Max Planck Institute for Astronomy, Heidelberg and the Max Planck Institute for Extraterrestrial Physics, Garching, The Johns Hopkins University, Durham University, the University of Edinburgh, the Queen's University Belfast, the Harvard-Smithsonian Center for Astrophysics, the Las Cumbres Observatory Global Telescope Network Incorporated, the National Central University of Taiwan, STScI, NASA under grant NNX08AR22G issued through the Planetary Science Division of the NASA Science Mission Directorate, National Science Foundation (NSF) grant AST-1238877, the University of Maryland, Eotvos Lorand University (ELTE), the Los Alamos National Laboratory, and the Gordon and Betty Moore Foundation. D.B. acknowledges support for this work provided by NASA through the NASA Hubble Fellowship grant HST-HF2-51430.001 awarded by STScI. 
A.V.F.'s group at U.C. Berkeley acknowledges generous support from Marc J. Staley (whose fellowship partly funded B.E.S. whilst contributing to the work presented herein as a graduate student), the TABASGO Foundation, the Christopher R. Redlich fund, the Miller Institute for Basic Research in Science (in which A.V.F. is a Miller Senior Fellow), and many individual donors. 
The UCSC team is supported in part by NSF grants AST-1518052 and AST-1815935, the Gordon \& Betty Moore Foundation, the Heising-Simons Foundation, and from fellowships from the Alfred P.\ Sloan Foundation and the David and Lucile Packard Foundation to R.J.F. D.A.C. acknowledges support from the National Science Foundation Graduate Research Fellowship under Grant DGE1339067. M.R.S. is supported by the National Science Foundation Graduate Research Fellowship Program under grant No. 1842400.

KAIT (for LOSS) and its ongoing operation were made possible by donations from Sun Microsystems, Inc., the Hewlett-Packard Company, AutoScope Corporation, Lick Observatory, the NSF, the University of California, the Sylvia \& Jim Katzman Foundation, and the TABASGO Foundation. Research at Lick Observatory is partially supported by a generous gift from Google.

Simulations, light-curve fitting, BBC, and cosmology pipeline are managed by \texttt{PIPPIN} \citep{Hinton_and_Brout2020}. Contours and parameter constraints are generated using the \textsc{ChainConsumer} package \citep{Hinton16}. Plots generated with Matplotlib \citep{matplotlib}. Usage of astropy \citep{astropy}, SciPy \citep{scipy}, and NumPy \citep{numpy}.

D.B. thanks his spouse Isabella and their future daughter for their support as the due date is rapidly approaching!

\appendix

\section{Data-Release Structure}
The data release will be found at \url{https://pantheonplussh0es.github.io/}.  It will also be found in the public version of SNANA in the directory ``Pantheon+''.  The link for the SNANA full data download is \url{https://zenodo.org/record/4015325}, and the SNANA source directory is  \url{https://github.com/RickKessler/SNANA}.

\begin{figure*}[htp]
    \centering

    \textbf{Light-curve file for SN 2021pit}\par\medskip

\includegraphics[clip,width=0.6\columnwidth]{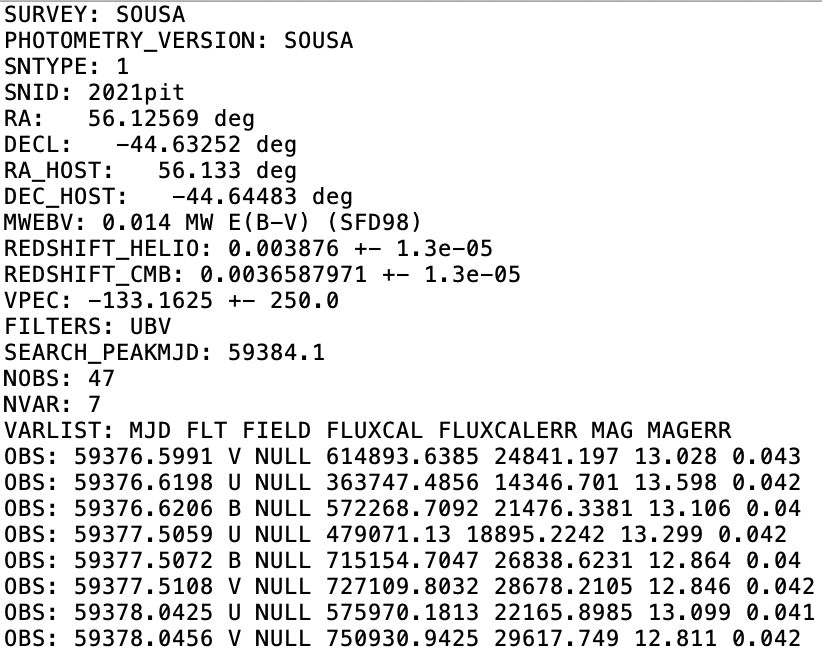}%

  \caption{Display of what an SNANA light-curve file looks like for SN~2021pit.  The full file is included at \url{https://pantheonplussh0es.github.io/}.}
      
    \label{fig:screenlc} \vspace{-1mm}
\end{figure*}

The structure of the Pantheon+ directory contains 18 subdirectories, with folder names synced to the data samples listed in Table~\ref{tab:surveys}.  In each folder, there is a .README file with documentation of the source of the data files, and a .LIST file which lists the SNe~Ia fitted as part of this analysis.  In each subdirectory, there are .txt files or .FITS describing the SN light curves along with meta-information.  If there are .txt files, the meta-information is at the top of the file, whereas if there is a .FITS file, the meta-information is in the HEAD.FITS file while the light-curve data are in the PHOT.FITS file.  For a single .txt file for one light curve, we show an example screenshot in Figure~\ref{fig:screenlc}.

The meta-information includes the following:
\begin{itemize}
    \item SN name
    \item SN position: RA, DEC in degrees and host position.
    \item SN Milky Way extinction (though this is overridden in fitting to ensure consistency with \cite{Schlafly11}.
    \item The heliocentric redshift, CMB-frame redshift, and peculiar velocity \textbf{VPEC}.
\end{itemize}
The light-curve data has columns for Date (\textbf{MJD}), Filter (\textbf{FLT}), Flux (\textbf{FLUXCAL}), Flux Uncertainty (\textbf{FLUXCALERR}), Magnitude (\textbf{MAG}), and Magnitude Uncertainty (\textbf{MAGERR}).  The common zero-point for all flux measurements is 27.5~mag.

We also include several global files for various properties.  These include the following:
\begin{itemize}
    \item List of Heliocentric, CMB, and Peculiar Velocities as derived by C22.
    \item List of all host-galaxy properties determined for this analysis.  These include mass for all SNe, and SFR and morphology for SNe with $z<0.15$.
\end{itemize}

\begin{figure*}[htp]
    \centering

    \textbf{FITRES file from SNANA's SALT2 light-curve fitting}\par\medskip

\includegraphics[clip,width=\columnwidth]{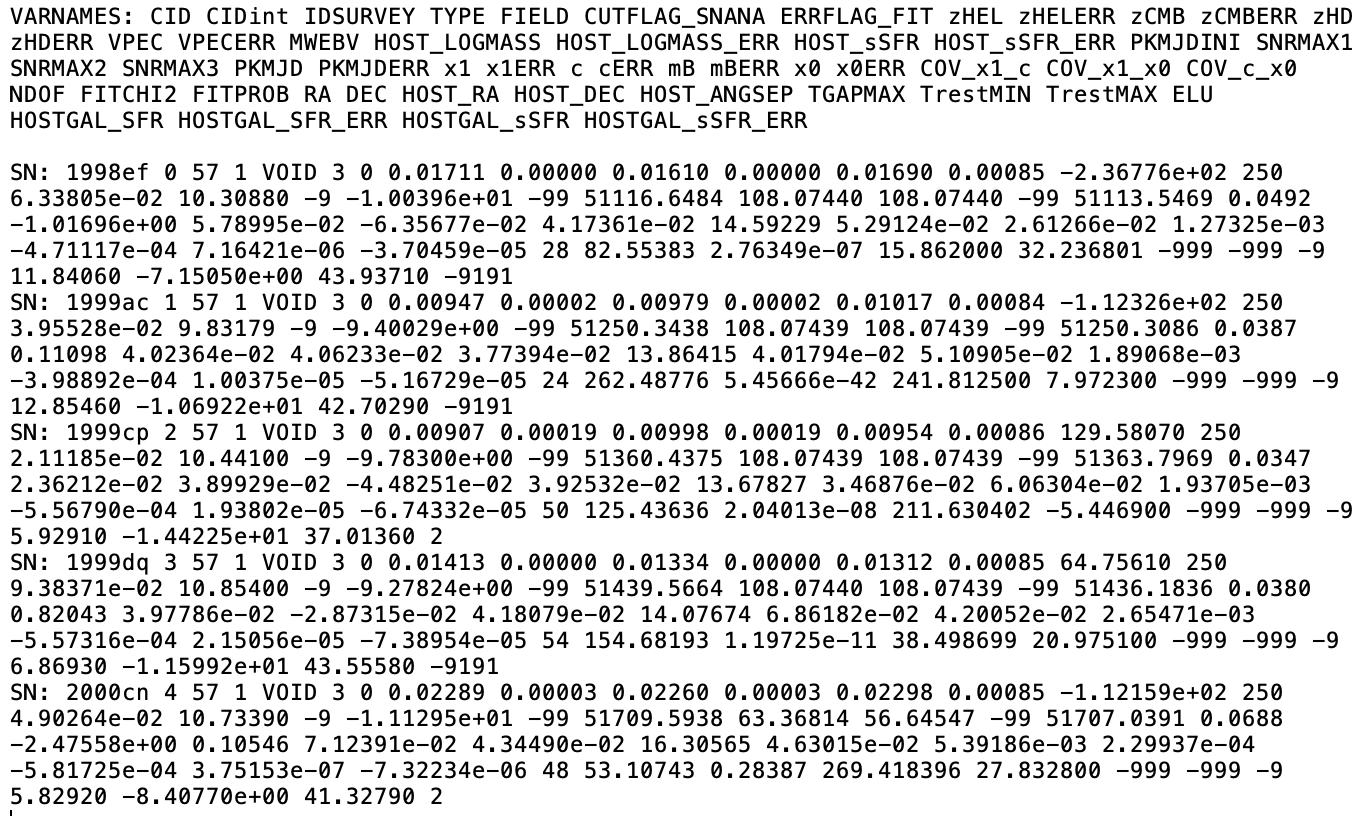}%

  \caption{Display of what a .FITRES file looks like that has all the information from the light-curve fit, as well as ancillary information.  A value of $-9$ is given where information is unavailable.  The full file will be included at \url{pantheonplussh0es.github.io}}
      
    \label{fig:input_fit} \vspace{-1mm}
\end{figure*}

Finally, we include a combined file of all the fitted parameters for each SN, before and after light-curve cuts are applied.  This is in the format of a .FITRES file and has all the meta-information listed above along with the fitted SALT2 parameters.  We show a screenshot of the release in Figure~\ref{fig:input_fit}.  Here, we give brief descriptions of each column.
\textbf{CID} -- name of SN.
\textbf{CIDint} -- counter of SNe in the sample.
\textbf{IDSURVEY} -- ID of the survey.
\textbf{TYPE} -- whether SN~Ia or not -- all SNe in this sample are SNe~Ia.
\textbf{FIELD} -- if observed in a particular field.
\textbf{CUTFLAG\_SNANA} -- any bits in light-curve fit flagged.
\textbf{ERRFLAG\_FIT} -- flag in fit.
\textbf{zHEL} -- heliocentric redshift.
\textbf{zHELERR} -- heliocentric redshift error.
\textbf{zCMB} -- CMB redshift.
\textbf{zCMBERR} -- CMB redshift error.
\textbf{zHD} -- Hubble Diagram redshift.
\textbf{zHDERR} -- Hubble Diagram redshift error.
\textbf{VPEC} -- peculiar velocity.
\textbf{VPECERR} -- peculiar-velocity error.
\textbf{MWEBV} -- MW extinction.
\textbf{HOST\_LOGMASS} -- mass of host.
\textbf{HOST\_LOGMASS\_ERR} -- error in mass of host.
\textbf{HOST\_sSFR} -- sSFR of host.
\textbf{HOST\_sSFR\_ERR} -- error in sSFR of host. 
\textbf{PKMJDINI} -- initial guess for PKMJD.
\textbf{SNRMAX1} -- First highest signal-to-noise ratio (SNR) of light curve.
\textbf{SNRMAX2} -- Second highest SNR of light curve.
\textbf{SNRMAX3} -- Third highest SNR of light curve.
\textbf{PKMJD} -- Fitted PKMJD.
\textbf{PKMJDERR} -- Fitted PKMJD error.
\textbf{x1} -- Fitted $x_1$.
\textbf{x1ERR} -- Fitted $x_1$ error.
\textbf{c} -- Fitted $c$.
\textbf{cERR} -- Fitted $c$ error.
\textbf{mB} -- Fitted $m_B$.
\textbf{mBERR} -- Fitted $m_B$ error.
\textbf{x0} -- Fitted $x_0$.
\textbf{x0ERR} -- Fitted $x_0$ error.
\textbf{COV\_x1\_c} -- covariance between $x_1$ and $c$.
\textbf{COV\_x1\_x0} -- covariance between $x_1$ and $x_0$.
\textbf{COV\_c\_x0} -- covariance between $c$ and $x_0$.
\textbf{NDOF} -- number of degrees of freedom (epochs) in light-curve fit.
\textbf{FITCHI2} -- $\chi ^2$ of light-curve fit.
\textbf{FITPROB} -- fit probability.
\textbf{RA} -- RA of SN (deg).
\textbf{DEC} -- DEC of SN (deg).
\textbf{HOST\_RA} -- RA of host (deg).
\textbf{HOST\_DEC} -- DEC of host (deg). 
\textbf{HOST\_ANGSEP} -- Separation between SN and host (deg).
\textbf{TGAPMAX} -- largest gap (in days) between observations.
\textbf{TrestMIN} -- minimum epoch with SN observation (day).
\textbf{TrestMAX} -- maximum epoch with SN observation (day).
\textbf{ELU} -- morphological classification.

\section{SN Data Information}
\label{sec:datainfo}

\begin{deluxetable*}{lr}
\tablecaption{Low-redshift SN photometry data releases and access}
\tablehead{\colhead{Question} & \colhead{Answer} \label{tab:information1}}
\startdata
CfA1&~\\
Where are the SN data published? & \citet{Riess99} \\
Where is the site for the SN data? & \url{https://www.cfa.harvard.edu/supernova/SNarchive.html} \\
What system is the SN in?	& Standard \\
\hline
CfA2&~\\
Where are the SN data published? & \citet{jha06} \\
Where is the site for the SN data? & \url{https://iopscience.iop.org/article/10.1086/497989/fulltext/204512.tables.html} \\
What system is the SN in?	& Standard \\
\hline
CfA3-Keplercam&~\\
Where are the SN data published? & \citet{Hicken2009} \\
Where is the site for the SN data? & \url{https://www.cfa.harvard.edu/supernova/CfA3/} \\
What system is the SN in?	& Standard and Natural \\
\hline
CfA3-4Shooter&~\\
Where are the SN data published? & \citet{Hicken2009} \\
Where is the site for the SN data? & \url{https://www.cfa.harvard.edu/supernova/CfA3/} \\
What system is the SN in?	& Standard and Natural \\
\hline
CfA4p1&~\\
Where are the SN data published? & \citet{Hicken2012} \\
Where is the site for the SN data? & \url{https://www.cfa.harvard.edu/supernova/CfA4/} \\
What system is the SN in?	& Standard and Natural \\
\hline
CfA4p2&~\\
Where are the SN data published? & \citet{Hicken2012} \\
Where is the site for the SN data? & \url{https://www.cfa.harvard.edu/supernova/CfA4/} \\
What system is the SN in?	& Standard and Natural \\
\hline
CNIa0.02&~\\
Where are the SN data published? & \cite{Chen20} \\
Where is the site for the SN data? & Private Communication \\
What system is the SN in?	& Natural \\
\hline
CSP DR3&~\\
Where are the SN data published? & \cite{Krisciunas2017} \\
Where is the site for the SN data? & \url{https://csp.obs.carnegiescience.edu/data/CSP_Photometry_DR3.tgz/view} \\
What system is the SN in?	& Natural \\
\hline
LOSS1&~\\
Where are the SN data published? & \cite{Ganeshalingam2010} \\
Where is the site for the SN data? & \url{http://heracles.astro.berkeley.edu/sndb/info} \\
What system is the SN in?	& Natural \\
\hline
LOSS2&~\\
Where are the SN data published? & \cite{Stahl2019} \\
Where is the site for the SN data? & \url{http://heracles.astro.berkeley.edu/sndb/info} \\
What system is the SN in?	& Natural \\
\hline
SOUSA&~\\
Where are the SN data published? & \cite{sousa14} \\
Where is the site for the SN data? & \url{https://pbrown801.github.io/SOUSA/} \\
What system is the SN in?	& VEGA \\
\hline
{Foundation}&~\\
Where are the SN data published? & \cite{Foley2018} \\
Where is the site for the SN data? & \url{https://github.com/djones1040/Foundation_DR1}\\
What system is the SN in?	& AB \\ 
\enddata
\tablecomments{In addition to the surveys listed here, there are individual releases of SN photometry as listed for LOWZ in Table \ref{tab:surveys}.}
\end{deluxetable*}

\begin{deluxetable*}{lr}
\tablecaption{High-redshift SN photometry data releases and access}
\tablehead{\colhead{Question} & \colhead{Answer} \label{tab:information2}}
\startdata
{PS1}&~\\
Where are the SN data published? & \cite{Scolnic2018} \\
Where is the site for the SN data? & \url{https://archive.stsci.edu/hlsps/ps1cosmo/jones/lightcurves/}\\
What system is the SN in?	& AB \\
\hline
{SDSS}&~\\
Where are the SN data published? & \cite{Sako2018} \\
Where is the site for the SN data? & \url{http://sdssdp62.fnal.gov/sdsssn/DataRelease/index.html}\\
What system is the SN in?	& AB \\
\hline
{SNLS}&~\\
Where are the SN data published? & \cite{Betoule2014} \\
Where is the site for the SN data? & \url{https://supernovae.in2p3.fr/sdss_snls_jla/ReadMe.html/}\\
What system is the SN in?	& AB \\
\hline
{DES}&~\\
Where are the SN data published? & \cite{Brout19} \\
Where is the site for the SN data? & \url{https://des.ncsa.illinois.edu/releases/sn/}\\
What system is the SN in?	& AB \\
\enddata
\tablecomments{In addition to the surveys listed here, there are \textit{HST} survey light curves as per the HDFN, SCP, CANDELS+CLASH, and GOODS+PANS references in Table~\ref{tab:surveys}.}
\end{deluxetable*}

\end{document}